\documentclass[aps,reprint,superscriptaddress,notitlepage,nofootinbib]{revtex4-1}

\usepackage{amssymb}
\usepackage{amsmath}
\usepackage{amsfonts}
\usepackage{amssymb}
\usepackage{graphicx}
\usepackage{subfigure}
\usepackage{url}
\usepackage{tikz}
\usepackage{amsthm}
\usepackage{algorithm}
\usepackage{algorithmic}

\usepackage[disable=true, textsize=footnotesize]{todonotes}
\usepackage[parfill]{parskip}
\usepackage{hyperref}
\usepackage{tabularx}

\usepackage{dsfont}

\usepackage{titlesec}

\usepackage{subfigure}
\usepackage{url}
\usepackage{algorithm}
\usepackage{algorithmic}

\usepackage{tabularx}

\newcommand{\dist}{\Delta}

\newcommand{\mat}[1]{\mathbf{#1}}

\newcommand{\Pref}{\widetilde{\mathrm{P}}_{st}^{\mathrm{ref}}}

\titlespacing{\section}{0pt}{8pt}{8pt}
\titlespacing{\subsection}{0pt}{8pt}{8pt}

\begin{abstract}
This paper introduces two new closely related betweenness centrality measures based on the Randomized Shortest Paths (RSP) framework, which fill a gap between traditional network centrality measures based on shortest paths and more recent methods considering random walks or current flows. 
The framework defines Boltzmann probability distributions over paths of the network which focus on the shortest paths, but also take into account longer paths depending on an inverse temperature parameter.
RSP's have previously proven to be useful in defining distance measures on networks.
In this work we study their utility in quantifying the importance of the nodes of a network.
The proposed RSP betweenness centralities combine, in an optimal way, the ideas of using the shortest and purely random paths for analysing the roles of network nodes, avoiding issues involving these two paradigms.
We present the derivations of these measures and how they can be computed in an efficient way.
In addition, we show with real world examples the potential of the RSP betweenness centralities in identifying interesting nodes of a network that more traditional methods might fail to notice.
\end{abstract}

\begin{document}

\title{Two betweenness centrality measures based on Randomized Shortest Paths}

\author{Ilkka Kivim\"aki}
\email[Corresponding author: ]{ilkka.kivimaki@aalto.fi}
\affiliation{Universit\'e Catholique de Louvain, ICTEAM/LSM, Place des Doyens 1, 1348 Louvain-la-Neuve, Belgium}
\affiliation{Aalto University, Department of Computer Science,\\ P.O.Box 15400, FI-00076 Aalto, Finland}

\author{Bertrand Lebichot}
\affiliation{Universit\'e Catholique de Louvain, ICTEAM/LSM, Place des Doyens 1, 1348 Louvain-la-Neuve, Belgium}

\author{Jari Saram\"aki}
\affiliation{Aalto University, Department of Computer Science,\\ P.O.Box 15400, FI-00076 Aalto, Finland}

\author{Marco Saerens}
\affiliation{Universit\'e Catholique de Louvain, ICTEAM/LSM, Place des Doyens 1, 1348 Louvain-la-Neuve, Belgium}

\maketitle

\vspace{-1cm}

\section{Introduction}\label{sec:Intro}

One of the most fundamental and popular topics in network science is determining the \textit{centrality} of a node in a network according to the structure of the network. 
The concept of centrality can be interpreted in many ways and a vast number of measures have been proposed based on different interpretations. 
One commonly used interpretation is \textit{betweenness centrality}, which reflects the extent to which a node lies in between pairs or groups of other nodes of the graph. 
This can be also stated as the extent to which a node is an intermediate in communication over the network. 
Different models have been proposed to measure the participation of a node in this communication ranging from the \textit{shortest path betweenness centrality} of Freeman \cite{Freeman-1977,Freeman-1978,Brandes-2001}, which considers communication flowing only along the shortest paths, to the \textit{current flow betweenness centrality} \cite{Newman-05, Brandes-2005b}, which interprets communication flowing as electric current or as random walks in the network.

In this article we propose two families of betweenness centrality measures based on the \textit{Randomized Shortest Paths} (RSP) framework \cite{RSP,Saerens-2008,kivimaki2014developments}. 
The framework is based on Boltzmann probability distributions over paths between the nodes of a network which focus on short, optimal paths, but give some probability mass also to longer paths. 
The extent of focus on optimal paths is controlled by an inverse temperature parameter $\beta$. 
The RSP framework has previously been shown to function well when defining distance measures on networks for clustering and classification of network nodes~\cite{kivimaki2014developments}. 
In this work we extend the study of RSP's by showing their potential also in defining centrality measures. 
The two RSP betweenness centralities presented in this paper measure the involvement of each node in RSP's between the nodes of the graph. 
The first measure, which we call the simple RSP betweenness, measures the expected number of visits to a node during RSP's, while the second one, called the RSP net betweenness is the sum of expected net flows over the edges connected to a node.

The proposed RSP betweenness measures are attractive both theoretically as well as in practice. 
Theoretical interest is ensured by the fact that both measures can be seen as generalizations of classical betweenness measures. 
With large values of the parameter $\beta$, both RSP betweenness measures converge to a measure that we introduce as the shortest path likelihood betweenness, which is very closely related to the original betweenness centrality defined by Freeman \cite{Freeman-1977,Freeman-1978,Brandes-2001} and its other similar variant, the load centrality \cite{Goh-2001}. 
The reason for defining two different betweenness measures with RSP's is in their behaviour as $\beta$ is decreased. 
Namely, the simple RSP betweenness then converges to the stationary distribution of a random walk on the network (multiplied by a constant), whereas the RSP net betweenness converges to the current flow betweenness \cite{Newman-05, Brandes-2005b}. 
The definition, as well as the computation, of the simple RSP betweenness are more straightforward than for its counterpart, the RSP net betweenness, which can also be stated of their corresponding limit functions. 
In addition, the experiments in Section \ref{sec:Exp} indicate that the simple RSP betweenness can in practice be more useful than the RSP net betweenness or the current flow betweenness. 
However, the choice of which definition to rely on in the end depends on the application domain.

Considering betweenness based on RSP's is motivated by the fact that measures based only on shortest paths or on random walks alone often involve undesirable features. 
Shortest paths in a complex network tend to pass through only a small fraction of the nodes of the network, which can cause highly skewed betweenness score distributions and fail in differentiating between the other nodes of the network \cite{Goh-2001,barthelemy2004betweenness}. 
Also, when considering communication or navigation in networks, it is not always even realistic to consider that they would occur along only the shortest, nor completely random paths. 
Instead, movement may follow a partly random route, with a drift towards a destination, for example when the navigating agent does not know the optimal way or wants to add secrecy and unpredictability to its route. 
As a result, the trajectories that are actually used in the network can be more spread over different nodes in areas of the network that contain many connections. 
Because of this, RSP's can help, for instance, in detecting bottlenecks of the network, where there exist no alternatives for the shortest path.

Measures based solely on random walks do take into account the abundance of connections between nodes. 
However, they may in many situations depend heavily on local features of a graph, especially for large graphs~\cite{Liben-Nowell-2007,nadler2009semi,vonLuxburg-2010} instead of capturing its interesting global properties. 
Thus, some regularisation over the degree of randomness is needed, which in the RSP framework is controlled by the inverse temperature parameter $\beta$.

Models that find a compromise between the optimal shortest path and a random walk have recently received a lot of attention. 
Compared to these, the attractive aspect in using the RSP framework is that the compromise between the shortest and random paths is optimal by definition, as the Boltzmann distribution minimizes the expected cost of paths subject to a fixed relative entropy~\cite{RSP, Saerens-2008}. 
The minimization can also be expressed with respect to free energy~\cite{kivimaki2014developments}. 
In addition to the optimality aspect, the computation of quantities related to RSP's is fairly straightforward and efficient. 
A drawback of the algorithms presented in the paper is that as such they are not tractable with very large networks. 
However, in the future we plan to develop more specialized methods that will enable the RSP quantities to be computed with large networks as well.

Ideas similar to the RSP framework of interpolating between the two extremes can be found in the work of Alamgir and von Luxburg \cite{pRes}, involving $p$-resistances and graph node distances based on them, of Chebotarev involving distances based on the matrix forest theorem \cite{Chebotarev-2011,Chebotarev-1997}, of Zhang and Boley \cite{ShortestToAllPath} with focus on routing schemes, of Estrada, who has defined the subgraph centrality \cite{estrada2005subgraph} and the communicability betweenness \cite{Estrada-2009}. 
Even more related to this paper are the recent works of Bavaud and Guex \cite{Bavaud-2012} and Lebichot et al.\ \cite{Lebichot-2014}. 
In fact, the two betweenness centrality measures presented in this paper can be shown equal to the betweenness measures proposed by Bavaud and Guex \cite{Bavaud-2012}, although the relation between the two works is not entirely obvious. 
We will discuss this relation and the relation between the RSP betweenness measures and other previously proposed betweenness measures in more detail in Section~\ref{sec:centralities}.

To sum up, the contributions of this paper are:
\begin{itemize}
\item We define two betweenness centrality measures which form a spectrum between measures based on shortest paths and pure random walks, therefore integrating information about both the optimality and the abundance of paths between nodes of the network,
\item We derive algorithms for computing these measures in a convenient and efficient way, and
\item We demonstrate with example networks that the proposed RSP betweenness measures may provide a more interesting ranking of the network nodes than the classical measures that they generalize.
\end{itemize}

The structure of the paper is as follows: Section \ref{sec:notations} introduces the notation and defines the terms as used in the paper. 
Section \ref{sec:centralities} lists existing network centrality measures with a focus on different interpretations of betweenness centrality. 
Section \ref{sec:RSPBC_section} reviews the RSP framework and then introduces the RSP betweenness centrality measures and the methodology for computing them. 
Section \ref{sec:Exp} presents example cases that illustrate the benefits of using RSP betweenness measures.

\section{Notation and terminology}
\label{sec:notations}

In the paper we consider weighted directed graphs $G=(V,E)$ with node set $V = \{1,2, \ldots , n\}$ and edge set $E = \{ (i,j) \}$ of $m$ edges. 
We define a path, or walk, interchangeably, 
as a sequence of nodes $\wp = (i_{0}, \ldots, i_{T})$, where $T > 0$ and $(i_{\tau},i_{\tau+1}) \in E$ for all $\tau = 0,\ldots,T-1$. 
A path is \textit{absorbing}, if the last node of the path appears on the path only once.
We denote the set of all absorbing paths starting from node $s$ and ending in node $t$, i.e.\ \textit{s-t-paths} or \textit{s-t-walks}, by $\mathcal{P}_{st}$.

The weights on edges, $a_{ij}, (i,j) \in E$, reflect the similarity or strength of connection between adjacent nodes and form the \textit{adjacency matrix} $\mathbf{A}$ of the graph. 
The edge weights define the \textit{reference transition probabilities} of the \textit{unbiased random walk} as $p_{ij}^{\mathrm{ref}} = a_{ij}/\sum_{j} a_{ij}$.
The transition probabilities form the \textit{reference transition probability matrix} $\mathbf{P}^{\mathrm{ref}}$, which can be computed as $\mathbf{P}^{\mathrm{ref}} = \mathbf{D}^{-1} \mathbf{A}$, where $\mathbf{D}$ is the diagonal matrix containing the row sums of $\mathbf{A}$.
The reference path probability $\Pref (\wp)$ of a path $\wp \in \mathcal{P}_{st}$ is simply defined as the product of the transition probabilities along the path.

In addition to the weights, the edges are also assigned costs $c_{ij}$, which, in contrast to weights, can be considered also as the dissimilarity or distance of adjacent nodes.
The cost of a path $\wp$ is then simply defined as $\tilde{c}(\wp) = \sum_{(i,j) \in \wp} c_{ij}$.
Accordingly, we use the term shortest path to mean the path between two nodes with the lowest cost over all paths between the nodes.
We denote the set of shortest paths from node $s$ to node $t$ by $\mathcal{P}^{*}_{st}$, the total number of such paths by $|\mathcal{P}^{*}_{st}|$ and the cost of the shortest path from $s$ to $t$ by $\tilde{c}_{st}^*$.
The directed graph that consists only of the nodes and directed edges that belong to the shortest paths from $s$ to $t$ is called the \textit{directed shortest path graph from s to t}.

In many situations the edge costs and edge weights can be defined based on one another, for example, as reciprocals $c_{ij} = 1/a_{ij}$, which corresponds to the interpretation of costs as resistances and weights as conductances in an electric circuit. 
This convention is also normally used when computing centrality measures on weighted networks. 
However, in general in the RSP framework the weights can be independent of the costs, as long as $c_{ij} < \infty$ whenever $a_{ij} > 0$. 
Accordingly, the transition probabilities and thus the unbiased random walk can be independent of the costs. 
This means that the edge costs define the interpretation of shortest paths, i.e.\ the low temperature behavior of the system, whereas the edge weights determine the interpretation of a random walk, i.e.\ the high temperature behavior.
The interplay between weights and costs is thus similar to a tradeoff between \textit{exploration}, based on local possible movements, and \textit{exploitation}, based on long-term preferred movements.

\section{Graph node centrality}
\label{sec:centralities}
The concept of graph node centrality has many interpretations, and for most of the interpretations there exists a lengthy catalog of different proposed measures, which are often derived for different application purposes. 
In addition there have been efforts of stating axioms that a centrality measure should have \cite{sabidussi1966centrality, Brandes-2005}. 
Also, as discussed by Kolaczyk \cite{Kolaczyk-2009}, there have been attempts to define a typology of centrality measures, for instance by Borgatti \cite{Borgatti-2006}.
In an interesting recent work, Brandes and Hildenbrand study the problem of finding minimal graphs for which different centrality measures rank different nodes as the most central~\cite{brandes2014smallest}.
In this section, we make a brief survey of the different centrality measures with special focus on betweenness centrality measures.

There are different possibilities in whether or not the source and target nodes, $s$ and $t$ should be considered also as intermediate nodes of a path for centrality considerations. 
In this paper we use the convention where the first and last node of a path do affect the betweenness scores. 
For betweenness measures based on shortest paths this choice only changes the overall betweenness scores by an additive constant in a strongly connected graph, and thus only affects the ranking of nodes when the network has several components. 
In contrast, when measuring betweenness based on random walks, further visits to the starting node $s$ after the first step may increase the betweenness score of node $s$ and thus may also affect the rankings. 
Betweenness measures are also often normalized, e.g.,\ according to the number of possible node pairs, when considering shortest paths between nodes. 
However, in this paper we leave the normalization out of consideration in all the definitions, because it never affects the rankings of nodes within a strongly connected network.

\subsection{Betweenness centralities}
\label{sec:betweenness_centralities}
\subsubsection{Betweenness based on shortest paths}
\label{sec:betweenness_centralities_sp}
Possibly the best-known centrality measure of all is the original \emph{betweenness centrality} of Freeman \cite{Freeman-1977,Freeman-1978,Brandes-2001}, which counts the fraction of shortest paths between a pair of nodes that an intermediate node lies on and sums these fractions over all node pairs. 
We will also refer to it as \emph{shortest path betweenness}, for specificity. 
Formally, the shortest path betweenness centrality of node $i$ can be expressed as
\begin{equation}
C_{i} = \sum_{s,t=1}^{n} \frac{n(i \in \mathcal{P}^{*}_{st})}{|\mathcal{P}^{*}_{st}|}, \label{eq:freeman_betweenness01} 
\end{equation}
where $n(i \in \mathcal{P}^{*}_{st})$ means the number of paths that contain node $i$.
Notice that if there are more than one shortest path connecting $s$ to $t$, each of these paths will contribute a score of $1/|\mathcal{P}^{*}_{st}|$ to the betweenness of the nodes on them.

There are several variations of the shortest path betweenness defined above. 
A thorough review of these variants and their efficient computation was provided by Brandes \cite{brandes2008variants}. 
One variant, called the \emph{load centrality}~\cite{Goh-2001,Newman2006Erratum,brandes2008variants}, replaces the fractional term inside the sum in (\ref{eq:freeman_betweenness01}) with the \textit{branching probability} of a path, i.e.\ the probability that a random walker moving in the directed shortest path graph from $s$ to $t$ follows a path that contains node $i$, when at each branching point in this graph it selects the edge to follow with uniform probability over all outgoing edges. 
For illustration of the difference between the shortest path betweenness and load betweenness, see \cite{brandes2008variants}. 
With weighted graphs the two measures are usually equal, as there often exists a unique shortest path between all nodes, especially if the weights are real-valued.

In what follows, we will need to consider another variant of the shortest path betweenness, which we have not encountered in the literature previously.
We call this measure the \textit{shortest path likelihood betweenness}.
Similarly to the load centrality, we define the likelihood betweenness by replacing the term inside the sum in (\ref{eq:freeman_betweenness01}) with the \textit{normalized likelihood} of a shortest path containing node $i$.
The \textit{likelihood} of a shortest path $\wp^* \in \mathcal{P}_{st}$ is the product of the reference transition probabilities along that path, i.e.\ the same as its reference path probability $\Pref(\wp^*)$.
Note, that the likelihood is different from the branching probability which is based on transition probabilities in the directed shortest path graph from $s$ to $t$, instead of the whole graph.
The \textit{normalized likelihood} of $\wp^*$ is then obtained by normalizing by the sum of likelihoods of all shortest paths, $\Pref(\wp^*) / \sum_{\wp \in \mathcal{P}_{st}^*} \Pref(\wp)$, which is also the contribution of $\wp^*$ to the shortest path likelihood betweenness of all the nodes along it.
The shortest path likelihood betweenness is introduced here for the sake of completeness because it is the limiting function of both RSP betweenness measures presented in this paper.
However, the differences between the three variants of shortest path betweenness presented above are very small, and in practice they provide very similar rankings of the nodes of a network.

\subsubsection{Betweenness based on random walks}
\label{sec:betweenness_rw}
Betweenness has also been considered with respect to other than just shortest paths, namely by considering random walks or flows on a network. 
The first such measure was proposed by Freeman \cite{Freeman-1991}, who defined the \emph{flow betweenness centrality} as the amount of flow through a node over maximum flows between all node pairs. 
The idea of considering flows was developed further by Newman \cite{Newman-05}, who defined the \emph{current flow betweenness centrality}, which measures the centrality of a node as the total sum of electrical current that flows through it, when considering all node pairs as source-sink pairs of a unit current flow. 
The current flow betweenness was also coined the \emph{random walk betweenness centrality} because of the well-established connection between electric current flows and random walks~\cite{Snell-1984}. 
Indeed, it can also be interpreted as the sum of expected net flows of a random walk over the edges connected to a node, meaning that the number of times that the walk enters or leaves the node along an edge cancel each other out.  Brandes and Fleischer \cite{Brandes-2005b} developed an algorithm which improves the efficiency of computing the current flow betweenness for all nodes of a network. 
The properties and computation of the current flow betweenness have also been studied by Bozzo and Franceschet \cite{bozzo2013resistance}.

Instead of considering net flows, a more straightforward definition of betweenness based on random walks is simply the overall expected number of visits to a node during a random walk. 
For arbitrary, non-absorbing walks, this quantity is not well-defined. 
However, it is possible to compute the proportion of steps of such a walk that the walker spends in a node. 
This is equal to the probability of finding the walker at the node after a long walk, i.e.\ the stationary probability of the node. 
The stationary probabilities define the \textit{stationary distribution}, which is the unique vector $\mathbf{\pi}$ that satisfies the equation $\mathbf{\pi} = (\mathbf{P}^{\mathrm{ref}})^{\mathsf{T}} \mathbf{\pi}$, given that the network is strongly connected and aperiodic. 
It is well known that for such graphs the stationary probabilities are proportional to the recurrence times and, if the graph is undirected, to the degree centralities (or strength, for weighted graphs) \cite{grinstead1997introduction}. 
For other graphs, the stationary distribution is not necessarily unique or may not even exist. 
To overcome this issue, for instance, the PageRank algorithm~\cite{Page-1998} uses a teleportation probability which transforms any kind of a graph into a strongly connected approximation and computes the stationary distribution on the transformed graph.

Even though the expected number of visits to a node is not well-defined for arbitrary walks, it is well-defined for absorbing walks. 
A bit surprisingly, we have discovered that when computed over all absorbing random walks on a strongly connected aperiodic directed graph, this measure is equal to the stationary distribution, up to a multiplying constant. 
The multiplying constant is the sum, over all $s$-$t$-pairs of the graph, of average hitting times, i.e.\ the expected number of steps along an absorbing path from $s$ to $t$, denoted by $\left< \tilde{c} \right>_{st}$. 
This result is not completely obvious and we have not found it mentioned explicitly in the literature. 
The proof of the result is omitted from the paper, as it is not relevant for this work, although the result itself is relevant for considering the RSP-based betweenness measures.

The RSP betweenness centrality measures proposed in this paper are also based on random walks, more particularly on the RSP framework, which we will review in more detail in Section~\ref{sec:RSP}. 
The first betweenness measure is based on counting the expected number of passages through a node of a random walker moving according to the RSP probabilities, while the other computes the expected net flow of walkers going through the node.

In fact, the RSP betweenness measures coincide with betweenness measures proposed recently, independently of our work, by Bavaud and Guex \cite{Bavaud-2012}. 
They propose a framework which interpolates between shortest paths and random walks by the minimization of free energy in a similar fashion as the free energy derivation of the RSP framework of Kivim\"aki et al.\ \cite{kivimaki2014developments}. 
One main difference in the work of Bavaud and Guex \cite{Bavaud-2012}, compared to the derivation of RSP's, is that a more general form of energy functionals, besides the expected length (or cost) of a path, is considered. 
In addition to that the relative entropy is considered with respect to transition probabilities instead of path probabilities, as is done in the definition of RSP's. 
The path distribution derived by Bavaud and Guex \cite{Bavaud-2012} can, however, be shown to equal the path distribution defined in the RSP framework, although this requires some lengthy and uninteresting derivations and is left out of this paper. 
The relation between the two approaches has also been studied by in the more recent work of Guex and Bavaud \cite{guex2015flow}. 
Compared to the work of Bavaud and Guex \cite{Bavaud-2012}, our work focuses more on the computational and practical aspects of the methodology. 
Thanks to the recent developments in the RSP framework \cite{kivimaki2014developments}, we are also able to present efficient algorithms for computing the quantities in question and illustrate the use of the methodology with practical examples.

Also closely related to the idea behind the RSP betweenness centralities is the \textit{bag-of-paths (BoP) betweenness centrality} \cite{Lebichot-2014}. 
It is based on the BoP framework, which also defines a Boltzmann distribution on the paths of a graph in a similar way as the RSP framework. 
The BoP betweenness is then defined as the a posteriori probability that a path selected according to the Boltzmann distribution visits an intermediate node $i$, when it walks from node $s$ to node $t$ according to the BoP probabilities. 
The BoP betweenness is also defined for groups and used for semi-supervised classification of graph nodes. 
A similar group betweenness measure can be defined from the RSP betweenness measures proposed here, but this is left out of the scope of this paper. 
The node classification task was also tackled by Devooght et al.\ \cite{devooght2014random} by using a modularity measure derived from the BoP framework.

\subsection{Other centralities}
\label{sec:Other_centralities}
Besides betweenness, centrality has also been characterized with other additional terms such as \emph{closeness}, \emph{feedback} and \emph{vitality}, to name a few~\cite{koschtzki2005centrality}. 
There have also been many efforts for stating axioms that would define centrality starting from the work of Sabidussi \cite{sabidussi1966centrality} and recently by Boldi and Vigna \cite{Boldi2014Axioms}. 
The concept of centrality can also be considered with respect to edges, instead of nodes. 
Although in this work we focus on centrality according to the betweenness interpretation, we nonetheless mention some of the other interpretations in this section, as the RSP framework could be combined with many of them too.

Closeness centrality measures are based on some interpretation of overall proximity of a node to the other nodes of a network. 
The \emph{shortest path closeness centrality} of node $i$ is classically defined as 
$C_i=1/ \sum_{j=1}^{n} \dist_{ij}$,

where $\dist_{ij}$ is the shortest path distance between nodes $i$ and $j$~\cite{bavelas1950communication, sabidussi1966centrality}. 
Considering communication on a network, betweenness centrality can be interpreted as the amount of \emph{control} of a node, whereas closeness centrality measures the \emph{efficiency} of the communication of the node~\cite{Freeman-1978,tutzauer2007entropy}. 
Instead of using the shortest path distance 
to define closeness centrality, other distances can be used too. 
For instance, White and Smyth \cite{White-2003} define \textit{Markov centrality} by replacing the shortest path distance in the closeness centrality definition with the average hitting time of an unbiased random walker, i.e., essentially the (unsymmetrized) commute time distance. 
Related to that, Brandes and Fleischer \cite{Brandes-2005b} considered a current flow analogy of closeness centrality by replacing the distance between two nodes with their potential difference. 
They managed to show that the current flow closeness centrality is equivalent to the information centrality defined by Stephenson and Zelen \cite{Stephenson-1989}. 
The equivalence has been confirmed with another proof by Bozzo and Franseschet \cite{bozzo2013resistance}. 
One subject for future research will be to extend closeness centrality by replacing the shortest path distance with RSP-based distance measures.

Feedback centrality measures are in many ways related to random walk based centrality measures. 
The \textit{eigenvector centrality} \cite{bonacich1972factoring} is the archetype feedback measure and is based on the idea that the centrality of a node should be the sum of the centralities of its neighbors. 
The solution of the formulation is the eigenvector of the adjacency matrix corresponding to the largest eigenvalue. 
The previously mentioned PageRank~\cite{Page-1998} can also be considered in this sense and formulated as an eigenvector. 
The Katz centrality~\cite{katz1953new} is based on the same idea of feedback, but considers also longer dependencies than only the neighbors of a node. 
The effect of longer dependencies decays according to the distance between nodes. 
The \textit{subgraph centrality} \cite{estrada2005subgraph} and \textit{communicability betweenness} \cite{Estrada-2009} are based on similar ideas using the matrix exponential of the adjacency matrix. 
The relationship between all the feedback centrality measures listed above has been studied by Benzi and Klymko \cite{benzi2013matrix}.

\section{Randomized shortest paths betweenness centralities}
\label{sec:RSPBC_section}
In this section we introduce two betweenness measures based on the randomized shortest paths (RSP) framework \cite{RSP,Saerens-2008,kivimaki2014developments}. 
These measures both generalize the shortest path likelihood betweenness measure defined in Section~\ref{sec:betweenness_centralities_sp}. 
In addition to that, one of the two measures also generalizes the stationary distribution of the graph while the other generalizes the current flow betweenness centrality. 
The RSP framework has previously been used for defining distance measures between graph nodes, which has proven useful for many data analysis tasks such as clustering and classification of graph nodes \cite{RSP,kivimaki2014developments}.

\subsection{Randomized shortest paths}
\label{sec:RSP}
In its core, the RSP framework \cite{RSP,Saerens-2008,Garcia-Diez-2011,kivimaki2014developments} is based on a probability distribution over paths between two nodes of a graph. 
The framework can also be formulated considering all paths, containing non-absorbing paths, but in this work, for simplicity, we restrict the framework to absorbing paths. 
The RSP probability distribution can be defined through either minimization of expected cost subject to a relative entropy constraint, or minimization of free energy \cite{kivimaki2014developments, Bavaud-2012}. 
Here we recall the former definition of minimization of expected cost, which is perhaps a more intuitive one.

The RSP framework is based on the probability distribution over the set $\mathcal{P}_{st}$ of absorbing $s$-$t$-walks for which the expected cost of the walks is minimal, when constrained with a fixed relative entropy with respect to the reference path probability distribution. 
Formally, we seek for the solution to the following problem
\begin{align}
\mathop{\text{Minimize}}_{\widetilde{\mathrm{P}}_{st}} \ 
&\sum_{\wp \in \mathcal{P}_{st}} \widetilde{\mathrm{P}}_{st}(\wp)\tilde{c}(\wp)
\nonumber
\\
\textrm{subject to} \ 
&\begin{cases}
J(\widetilde{\mathrm{P}}_{st} \| \Pref) = J_{0} \\ 
\sum\limits_{\wp \in \mathcal{P}_{st}}\widetilde{\mathrm{P}}_{st}(\wp) = 1
\end{cases}
\end{align}
where $\Pref(\wp)$ is the reference path probability, $\widetilde{c}(\wp)$ the overall cost of path $\wp$ and $J(\widetilde{\mathrm{P}}_{st} \| \Pref)$ is the relative entropy, or Kullback-Leibler divergence, which is set to a desired level $J_0$.

The solution of this minimization is a Boltzmann distribution (for details, see \cite{RSP, Saerens-2008}):
\begin{equation}
\label{eq:RSPprob}
\widetilde{\mathrm{P}}_{st}(\wp) = \dfrac{\Pref(\wp) \exp \left( -\beta \widetilde{c}(\wp)\right)}{\mathcal{Z}_{st}},
\end{equation}
where
\begin{equation}
\label{eq:partition_function}
\mathcal{Z}_{st} = \sum\limits_{\wp \in \mathcal{P}_{st}}\Pref(\wp) \exp \left( -\beta \widetilde{c}(\wp)\right)
\end{equation}
is the \textit{partition function} of absorbing walks from $s$ to $t$ and the inverse temperature parameter $\beta$ controls the divergence from the unbiased random walk probabilities. 
When applying the framework, the user is supposed to input the value for $\beta$, instead of the relative entropy $J_0$. 
Low and high values of $\beta$ correspond, respectively, to low and high values of $J_0$ and, inversely, to high and low temperature. 
In other words, for high values of $\beta$ (low temperature), the path distribution $\widetilde{\mathrm{P}}_{st}$ focuses on shortest paths, whereas for low values of $\beta$ (high temperature) more random paths are also preferred. 
It is also possible to find the model corresponding to a particular value of $J_0$, for instance, with a binary search over different values of $\beta$.

The partition function $\mathcal{Z}_{st}$ plays an important role in the derivation of the computation of many quantities related to the RSP framework, as can be seen in the following. 
Concerning the computation of $\mathcal{Z}_{st}$, we refer to earlier works of Fran\c{c}oisse et al.\ \cite{BoP} and Kivim\"aki et al.\ \cite{kivimaki2014developments} which show that it can be expressed as
\begin{equation}
\mathcal{Z}_{st} = \frac{z_{st}}{z_{tt}},
\label{eq:Partition_function_hitting01}
\end{equation}
where $z_{st}$ is the element $(s,t)$ of the \textit{fundamental matrix of non-absorbing paths}, defined as 
\begin{equation}
\mathbf{Z}
= (\mathbf{I} - \mathbf{W})^{-1}, \text{\quad with \quad} 
\mathbf{W} = \mathbf{P}^{\mathrm{ref}} \circ \exp(-\beta \mathbf{C}),
\end{equation}
where $\circ$ and $\exp$ denote the element-wise matrix multiplication and exponential, respectively. 
The matrix $\mathbf{W}$, defined from the reference transition probability matrix $\mathbf{P}^{\mathrm{ref}}$ and the cost matrix $\mathbf{C}$, is a substochastic matrix and can be interpreted as defining a \emph{killed random walk}. 
Consequently, the partition function $\mathcal{Z}_{st}$ can then be interpreted as the probability of a walker surviving the walk from $s$ to $t$ (see\cite{BoP,kivimaki2014developments} for details).

\subsection{The simple RSP betweenness}
\label{sec:RSPBC}
We first define the simple RSP betweenness centrality of a node $i$ \textit{with respect to absorbing paths from} $s$ \textit{to} $t$, as the expected number of visits through $i$ over all $s$-$t$-walks, denoted by $\overline{n}_i(s,t)$, with respect to the RSP probabilities of Equation~(\ref{eq:RSPprob}). 
This can further be expressed based on the expected number of passages through edges leaving from node $i$ over all $s$-$t$-walks, denoted by $\overline{\eta}_{ij}(s,t)$, as: 
\begin{equation}
\mathrm{bet}_{i}^{\text{RSP}}(s,t) = \overline{n}_i(s,t) = \sum_{j: (i,j) \in E} \overline{\eta}_{ij}(s,t)
\end{equation}
The function $\mathrm{bet}_{i}^{\text{RSP}}$ can be useful for visualizing path distributions and for path planning tasks between two fixed nodes of the graph \cite{Garcia-Diez-2011,Panzacchi-2015}. 
Moreover, it can be used to investigate how central an intermediate actor is with respect to, for instance, the communication between two other actors in a social network.

We then define the overall \textit{simple RSP betweenness centrality} of node $i$ as the sum of contributions over all source-target pairs, on the graph:
\begin{equation}
\mathrm{bet}_{i}^{\text{RSP}} = 
\sum_{s=1}^{n} \sum_{t=1}^{n} \mathrm{bet}_{i}^{\text{RSP}}(s,t).
\label{eq:randomized_shortest_path_betweenness_definition01}
\end{equation}
Next, we derive the method for computing this quantity in closed form.
Let us denote by $\eta(i \rightarrow j \in \wp)$ the number of times that the edge $(i,j)$ is traversed along the path $\wp$. 
Then, by writing out the expression of $\overline{\eta}_{ij}(s,t)$, the use of the partition function in the computation of the RSP quantities becomes evident:
\begin{align}
  \bar{\eta}_{ij}(s,t) &= \displaystyle \sum_{\wp\in\mathcal{P}}  \widetilde{\mathrm{P}}_{st}(\wp) \, \eta(i \rightarrow j \in \wp) \nonumber \\
  &= \frac{{\displaystyle \sum_{\wp\in\mathcal{P}}} \tilde{\mathrm{P}}^{\text{ref}}(\wp)\exp\left[-\beta \tilde{c}(\wp)\right] \, \eta(i \rightarrow j \in \wp)}{{\displaystyle \sum_{\wp'\in\mathcal{P}}}\tilde{\mathrm{P}}^{\text{ref}}(\wp')\exp\left[-\beta \tilde{c}(\wp')\right]} \nonumber
  \\ 
  &= \frac{{\displaystyle \sum_{\wp\in\mathcal{P}}} \tilde{\mathrm{P}}^{\text{ref}}(\wp)\exp\left[-\beta \tilde{c}(\wp)\right] \, \dfrac{
	\partial \tilde{c}(\wp)}{
	\partial c_{ij}}}{{\displaystyle \sum_{\wp'\in\mathcal{P}}}\tilde{\mathrm{P}}^{\text{ref}}(\wp')\exp\left[-\beta \tilde{c}(\wp')\right]} \nonumber \\
	&=	-\frac{1}{\beta} \dfrac{
	\partial \log \mathcal{Z}_{st}}{
	\partial c_{ij}} 
	\label{eq:Expected_visits_hitting01}
\end{align}
Thus, the expected number of transitions $i \rightarrow j$ over $s$-$t$-walks can be computed by differentiating the logarithm of the partition function, which is common knowledge in statistical physics (see, e.g.,\cite{Peliti-2011}). 
Note that (\ref{eq:Expected_visits_hitting01}) holds only if there exists a path from $s$ to $t$. 
Otherwise, naturally, $\bar{\eta}_{ij}(s,t) = 0$.

By combining Equations (\ref{eq:Partition_function_hitting01}) and (\ref{eq:Expected_visits_hitting01}), the computation of $\bar{n}_{ij}(s,t)$ can be written as:
\begin{align}
	\label{eq:number_passages_link01} 
	\bar{\eta}_{ij}(s,t) 
 &= -\frac{1}{\beta} \dfrac{\partial \log \mathcal{Z}_{st}}{\partial c_{ij}}
	= -\frac{1}{\beta} \dfrac{\partial \log (z_{st}/z_{tt})}{\partial c_{ij}} 
  \nonumber \\
	&= -\frac{1}{\beta} \left( \dfrac{1}{z_{st}} \dfrac{
	\partial z_{st}}{
	\partial c_{ij}} - \dfrac{1}{z_{tt}} \dfrac{
	\partial z_{tt}}{
	\partial c_{ij}} \right).
\end{align}
Therefore, we need to compute $\partial z_{st}/ \partial c_{ij}$, which can be achieved using matrix formalism. 
If we denote by $\mathbf{e}_i$ the ($n \times 1$)-vector whose element $i$ is 1 and others are 0, then 
\begin{align}
	&\dfrac{
	\partial z_{st}}{
	\partial c_{ij}} 
	= \dfrac{
	\partial (\mathbf{e}_{s}^{\mathsf{T}} \mathbf{Z} \mathbf{e}_{t})}{
	\partial c_{ij}}
	= \dfrac{
	\partial (\mathbf{e}_{s}^{\mathsf{T}} (\mathbf{I} - \mathbf{W})^{-1} \mathbf{e}_{t})}{
	\partial c_{ij}} \nonumber \\
	&= \mathbf{e}_{s}^{\mathsf{T}} \dfrac{
	\partial (\mathbf{I} - \mathbf{W})^{-1}}{
	\partial c_{ij}} \mathbf{e}_{t}
	= \mathbf{e}_{s}^{\mathsf{T}} \left( - \mathbf{Z} \dfrac{
	\partial (\mathbf{I} - \mathbf{W})}{
	\partial c_{ij}} \mathbf{Z} \right) \mathbf{e}_{t} \nonumber
	\\
	&= \mathbf{e}_{s}^{\mathsf{T}} \left( \mathbf{Z} \dfrac{
	\partial \mathbf{W}}{
	\partial c_{ij}} \mathbf{Z} \right) \mathbf{e}_{t} \nonumber 
	= - \beta \, p_{ij}^{\text{ref}} \exp[- \beta c_{ij}] \mathbf{e}_{s}^{\mathsf{T}} \mathbf{Z} \mathbf{e}_{i} \mathbf{e}_{j}^{\mathsf{T}} \mathbf{Z} \mathbf{e}_{t} \nonumber \\
	&= - \beta w_{ij} z_{si} z_{jt} 
\end{align}
where we used $\dfrac{
\partial \mathbf{X}^{-1}}{
\partial x} = - \mathbf{X}^{-1} \dfrac{
\partial \mathbf{X}}{
\partial x} \mathbf{X}^{-1}$ (see, e.g., \cite{Harville-1997,Seber-2008}).
Equation (\ref{eq:number_passages_link01}) can therefore be rewritten as 
\begin{align}
	\bar{\eta}_{ij}(s,t) 
	&=  \dfrac{1}{z_{st}} z_{si} w_{ij} z_{jt} - \dfrac{1}{z_{tt}} z_{ti} w_{ij} z_{jt}  
	\nonumber \\
	&= \left( \dfrac{z_{si}}{z_{st}} - \dfrac{z_{ti}}{z_{tt}} \right) w_{ij} z_{jt} \label{eq:number_passages_link02} 
\end{align}
Furthermore, the total flow transiting through node $i$, given that $i \neq t$, is 
\begin{equation}
	\bar{n}_{i}(s,t) = \sum_{j=1}^{n} \bar{\eta}_{ij}(s,t) = \left( \dfrac{z_{si}}{z_{st}} - \dfrac{z_{ti}}{z_{tt}} \right) \sum_{j=1}^{n} w_{ij} z_{jt}
	\label{eq:number_passages_node01} 
\end{equation}

The expression on the right-hand side of Equation (\ref{eq:number_passages_node01}) can be further simplified in the following way. 
We know that $(\mathbf{I} - \mathbf{W})^{-1} (\mathbf{I} - \mathbf{W}) = \mathbf{I}$, which implies that $\mathbf{Z} (\mathbf{I} - \mathbf{W}) = \mathbf{I}$ and therefore that $\mathbf{Z} = \mathbf{Z} \mathbf{W} + \mathbf{I}$, or element-wise, $z_{it} = \sum_{j=1}^{n} w_{ij} z_{jt} + \delta_{it}$, where $\delta_{it}$ is the Kronecker delta. 
However, the term $\delta_{it}$ can be discarded, when considering Equation~(\ref{eq:number_passages_node01}), because when $i=t$, we anyway have $(z_{si}/z_{st} - z_{ti}/z_{tt}) = 0$. 
Thus, Equation (\ref{eq:number_passages_node01}) simplifies to 
\begin{equation}
  \bar{n}_{i}(s,t) = \left( \dfrac{z_{si}}{z_{st}} - \dfrac{z_{ti}}{z_{tt}} \right) z_{it}
  \label{eq:number_passages_node02} 
\end{equation}

Finally, the simple RSP betweenness centrality (Equation~(\ref{eq:randomized_shortest_path_betweenness_definition01}))  can be computed with matrix manipulation as
\begin{align}
	&\mathrm{bet}_{i}^{\text{RSP}} = \sum_{s,t = 1}^{n} \bar{n}_{i}(s,t) = \sum_{s,t = 1}^{n} \left( \dfrac{z_{si}}{z_{st}} - \dfrac{z_{ti}}{z_{tt}} \right) z_{it} \nonumber \\
	&=  \sum_{s,t = 1}^{n} z_{it} \dfrac{1}{z_{ts}^{\text{t}}} z_{si} - \sum_{s,t = 1}^{n} z_{it} \dfrac{1}{z_{tt}} z_{ti} \nonumber \\
	&=  \sum_{s,t = 1}^{n} z_{it} \dfrac{1}{z_{ts}^{\text{t}}} z_{si} - n \sum_{t = 1}^{n} z_{it} \dfrac{1}{z_{tt}} z_{ti} \nonumber \\
	&= \left[ \mathbf{diag} \left( \mathbf{Z} \left( \mathbf{Z}^{\div} \right)^{\mathsf{T}} \mathbf{Z} \right) - n \, \mathbf{diag} \left(\mathbf{Z} \, \mathbf{Diag} \left( \mathbf{Z}^{\div} \right) \mathbf{Z} \right) \right]_{i} \nonumber \\
	&= \left[ \mathbf{diag} \left(\mathbf{Z} \left( \mathbf{Z}^{\div} - n \, \mathbf{Diag} \left( \mathbf{Z}^{\div}\right) \right)^{\mathsf{T}} \mathbf{Z} \right) \right]_{i} 
	\label{eq:delta_decomposition_term01} 
\end{align}
where $\mathbf{diag}(\mathbf{X})$ and $\mathbf{Diag}(\mathbf{X})$ are, respectively, a column vector and a diagonal matrix containing the diagonal of $\mathbf{X}$, $\mathbf{X}^{\div}$ denotes the element-wise reciprocal matrix, i.e., $x_{ij}^{\div} = 1/x_{ij}$ and the superscript $\text{t}$ denotes elements from the transposed matrix, i.e., $z_{ij}^{\text{t}} = z_{ji}$. 
The vector $\mathbf{bet}^{\mathrm{RSP}}$ of all betweenness values is computed accordingly.

The pseudocode for computing the simple RSP betweenness for all nodes is presented in Algorithm~\ref{alg:randomized_shortest_path_betweenness01}.\footnote{The Matlab code for the algorithms presented in the paper, as well as the real world networks used in the experiments, are available online at \href{https://github.com/ikivimak/RSP-betweenness}{https://github.com/ikivimak/RSP-betweenness}
} In conclusion, the simple RSP betweenness scores of all nodes can be computed by performing the matrix inversion $\mathbf{Z} = (\mathbf{I}-\mathbf{W})^{-1}$ and then simple matrix operations according to Equation~(\ref{eq:delta_decomposition_term01}). 
Thus, the computational bottleneck of the algorithm is the matrix inversion, which, in general, has time complexity $\mathcal{O}(n^3)$ and space complexity $\mathcal{O}(n^2)$, because of which the method is currently not practical with very large networks.

\begin{algorithm}[H]
\caption{\small{Computing the simple RSP betweenness vector of a graph $G$.}}

\algsetup{indent=1em, linenodelimiter=.}

\begin{algorithmic}[1]
\small
\REQUIRE $\,$ \\
 -- A directed strongly connected graph $G$ with $n$ nodes. \\
 -- The $n\times n$ reference transition probability matrix $\mathbf{P}^{\mathrm{ref}}$ (defined from the adjacency matrix as $\mathbf{P}^{\mathrm{ref}} = \mathbf{D}^{-1}\mathbf{A}$) \\ 
 -- The $n\times n$ non-negative cost matrix $\mathbf{C}$ \\
 -- The inverse temperature parameter $\beta$.\\
 
\ENSURE $\,$ \\
 -- The $n \times 1$ simple RSP betweenness vector $\mathbf{bet}^{\mathrm{RSP}}$.\\

\STATE $\mathbf{W} \leftarrow \mathbf{P}^{\mathrm{ref}} \circ \exp\left[-\beta\mathbf{C}\right]$ 
\STATE $\mathbf{Z} \leftarrow (\mathbf{I}-\mathbf{W}\mathbf{)}^{-1}$ 
\STATE $\mathbf{Z}^{\div} \leftarrow \mathbf{e} \mathbf{e}^{\mathsf{T}} \div \mathbf{Z}$ 
\STATE $\mathbf{bet}^{\mathrm{RSP}} \leftarrow \mathbf{diag} \left(\mathbf{Z} (\mathbf{Z}^{\div} - n \, \mathbf{Diag}(\mathbf{Z}^{\div}))^{\mathsf{T}} \mathbf{Z} \right) $ 

\RETURN $\mathbf{bet}^{\mathrm{RSP}}$

\end{algorithmic}
\label{alg:randomized_shortest_path_betweenness01} 
\end{algorithm}

\subsection{RSP net betweenness}
\label{sec:RSPNBC}
Instead of only considering the overall outgoing flow of random walkers, as in the definition of Equation (\ref{eq:randomized_shortest_path_betweenness_definition01}), it may in some cases make more sense to compute the net outgoing flow~\cite{RSP}, i.e.\ so that the outgoing and ingoing flows through one edge neutralize each other. 
This corresponds to the random walk interpretation of the current flow betweenness in undirected graphs~\cite{Newman-05,Brandes-2005b,bozzo2013resistance}. 
According to this approach, we define the \textit{RSP net betweenness centrality} of node $i$ as 
\begin{equation}
\mathrm{bet}_{i}^{\text{RSPnet}} = 
 \sum_{s=1}^{n} \sum_{t=1}^{n} \sum_{j: (i,j) \in E} |\bar{\eta}_{ij}(s,t) - \bar{\eta}_{ji}(s,t)|
 \label{eq:randomized_shortest_path_net_betweenness01} 
\end{equation}
The computation of this quantity for all nodes at once is a bit more involved than in the case of the simple RSP betweenness, because of the absolute value in the expression. 
A na\"ive algorithm would loop over all $s$-$t$-pairs and compute the contribution of the corresponding paths to each intermediate node. 
However, a more efficient solution is to perform a loop over each edge of the network, compute separately the net flow through that edge over all $s$-$t$-paths, and to add this net flow to the betweenness score of the starting node of the edge. 
This change in looping strategy improves the complexity by a factor from $\mathcal{O}(n^2)$ to $\mathcal{O}(m)$, which makes a big difference, when dealing with a sparse network. 
The faster computation can be achieved by writing out the matrix $\mathbf{N}^{ij}$, whose element $(s,t)$ is $\bar{\eta}_{ij}(s,t)$ from Equation (\ref{eq:number_passages_link02}):
\begin{align}
\label{eq:matrix_of_net_flows}
&\left[\mathbf{N}^{ij}\right]_{st} 
= \bar{\eta}_{ij}(s,t)
= \left( \dfrac{z_{si} z_{jt}}{z_{st}} - \dfrac{z_{ti} z_{jt}}{z_{tt}} \right) w_{ij} \nonumber \\
&= \left[ \left( \mathbf{z}_i^{\mathrm{c}} (\mathbf{z}_j^{\mathrm{r}})^{\mathsf{T}} \div \mathbf{Z} - \mathbf{e}\left(\mathbf{z}_i^{\mathrm{c}} \circ \mathbf{z}_j^{\mathrm{r}} \div \mathbf{diag}(\mathbf{Z}) \right)^{\mathsf{T}} \right) \right]_{st} \! w_{ij},
\end{align}
where $\circ$ and $\div$ denote elementwise matrix multiplication and division, respectively, $\mathbf{z}_i^{\mathrm{c}}$ and $\mathbf{z}_j^{\mathrm{r}}$ denote the $(n \times 1)$-vectors corresponding, respectively, to the $i$-th column and $j$-th row of matrix $\mathbf{Z}$ and $\mathbf{e}$ is the $(n \times 1)$-vector whose all elements are 1.

The overall net flow through edge $(i,j)$ can then be computed as
\begin{equation}
\label{eq:matrix_form_of_net_flow}
\bar{\eta}_{ij}^{\mathrm{net}}
= \sum_{s,t=1}^{n} \left[ \ |\mathbf{N}^{ij}-\mathbf{N}^{ji}| \ \right]_{st}
= \mathbf{e}^{\mathsf{T}} |\mathbf{N}^{ij}-\mathbf{N}^{ji}|\mathbf{e}
\end{equation}
after which the betweenness score of each node can be simply computed by summing up the contributions of each edge connected to the node:
\begin{equation}
\mathrm{bet}_{i}^{\mathrm{RSPnet}} = \sum_{j: (i,j) \in E} \bar{\eta}_{ij}^{\mathrm{net}}.
\end{equation}
Algorithm~\ref{alg:randomized_shortest_path_net_betweenness01} contains the pseudocode for computing the RSP net betweenness. 
In principle, the algorithm can be used with directed graphs and the result can be interpreted according to the net flow of random walks, even though the electric current interpretation only makes sense with undirected graphs. 
With undirected graphs, Algorithm~\ref{alg:randomized_shortest_path_net_betweenness01} should be altered so that it only considers each undirected edge only once and increments also the betweenness of node $j$ in addition to node $i$ at step 10. 
This reduces the computation time on undirected networks by a half. 
Algorithm~\ref{alg:randomized_shortest_path_net_betweenness01} also contains the same matrix inversion as Algorithm~\ref{alg:randomized_shortest_path_betweenness01} of time complexity $\mathcal{O}(n^3)$. 
In addition to this, the other consuming task is the loop over all $m$ edges of the graph in steps 4.-11., inside which elementary matrix operations of time complexity $\mathcal{O}(n^2)$ have to be performed. 
Thus, the total time complexity of the algorithm is $\mathcal{O}(n^3 + mn^2)$.

Although the RSP net betweenness can be computed for directed graphs, we have not found a good use case for this purpose. 
The definition on directed graphs is not as intuitive as the simple betweenness and, moreover, current flows and the current flow betweenness are normally defined only for undirected graphs. 
Nevertheless, it is possible to use Algorithm~\ref{alg:randomized_shortest_path_net_betweenness01} with directed graphs, in addition to which it is possible to derive a similar algorithm for computing the directed version of the current flow betweenness based on the pseudoinverse of the Laplacian matrix of the graph.

\begin{algorithm}[H]
\caption{\small{The RSP net betweenness vector of a graph $G$.}}

\algsetup{indent=1em, linenodelimiter=.}

\begin{algorithmic}[1]
\small
\REQUIRE $\,$ \\
Same as Algorithm~\ref{alg:randomized_shortest_path_betweenness01}. 
\ENSURE $\,$ \\
 -- The $n \times 1$ RSP net betweenness vector $\mathbf{bet}^{\mathrm{RSPnet}}$.\\
\STATE  $\mathbf{W} \leftarrow \mathbf{P}^{\mathrm{ref}}\circ\exp\left[-\beta\mathbf{C}\right]$ 
\STATE $\mathbf{Z} \leftarrow (\mathbf{I}-\mathbf{W}\mathbf{)}^{-1}$ 
\STATE $\mathbf{bet}^{\mathrm{RSPnet}} \leftarrow \mat{0}$ 
\FOR{$i=1$ to $n$}
\STATE $\mat{z}^{\mathrm{c}}_{i} \leftarrow \mathbf{Ze}_i$, $\mat{z}^{\mathrm{r}}_{i} \leftarrow \mathbf{Z}^{\mathsf{T}}\mathbf{e}_i$ 
\FORALL{$j$ such that $(i,j) \in E$}
\STATE $\mat{z}^{\mathrm{c}}_{j} \leftarrow \mathbf{Ze}_j$, $\mat{z}^{\mathrm{r}}_{j} \leftarrow \mathbf{Z}^{\mathsf{T}}\mathbf{e}_j$ 
\STATE $\mat{N}^{ij}
\leftarrow w_{ij} \! \left[ \left( \mat{z}^{\mathrm{c}}_{i} (\mat{z}^{\mathrm{r}}_{j})^{\mathsf{T}} \div \mat{Z} \right)
- \mat{e} \left( (\mat{z}^{\mathrm{c}}_{i} \circ \mat{z}^{\mathrm{r}}_{j}) \div \mat{diag}(\mat{Z}) \right)^{\mathsf{T}} \right]$ 
\STATE $\mat{N}^{ji}
\leftarrow w_{ji} \! \left[ \left( \mat{z}^{\mathrm{c}}_{j} (\mat{z}^{\mathrm{r}}_{i})^{\mathsf{T}} \div \mat{Z} \right)
- \mat{e} \left( (\mat{z}^{\mathrm{c}}_{j} \circ \mat{z}^{\mathrm{r}}_{i}) \div \mat{diag}(\mat{Z}) \right)^{\mathsf{T}} \right]$ 
\STATE $\mathrm{bet}_{i}^{\mathrm{RSPnet}} \leftarrow \mathrm{bet}_{i}^{\mathrm{RSPnet}} + \mat{e}^{\mathsf{T}} \left| \mat{N}^{ij} - \mat{N}^{ji} \right| \mat{e}$
\ENDFOR
\ENDFOR
\RETURN $\mathbf{bet}^{\mathrm{RSPnet}}$

\end{algorithmic}
\label{alg:randomized_shortest_path_net_betweenness01} 
\end{algorithm}

\subsection*{RSP betweenness centralities at the limit $\beta \longrightarrow \infty$}
The simple RSP betweenness centrality counts the expected number of visits to each node during RSP's between all source-target pairs of the graph. 
In the low temperature limit, i.e.\ when $\beta \to \infty$, the RSP probability distribution of Equation (\ref{eq:RSPprob}) focuses solely on the shortest paths of the graph and the expected number of visits to a node on a shortest path approaches the probability of following that particular path. 
Moreover, when $\beta \to \infty$, for all paths $\wp \in \mathcal{P}_{st}$, whose cost $\tilde{c}(\wp) > \tilde{c}_{st}^*$, $\exp(-\beta \tilde{c}(\wp)) \ll \exp(-\beta \tilde{c}_{st}^*)$, because of which the partition function $\mathcal{Z}_{st}$ of Equation (\ref{eq:partition_function}) becomes dominated by the terms determined by the shortest paths. 
As a result,
\begin{align}
\widetilde{\mathrm{P}}_{st}(\wp)
&= \dfrac{\Pref(\wp) \exp \left( -\beta \widetilde{c}(\wp)\right)}{\mathcal{Z}_{st}} 
\nonumber
\\
&\mathop{\longrightarrow}_{\beta \rightarrow \infty} 
\begin{cases}
0, & \text{if} \ \wp \notin \mathcal{P}_{st}^{*} \\
\dfrac{\Pref(\wp)}{\sum_{\wp \in \mathcal{P}_{st}^{*}} \Pref(\wp)}, & \text{if} \ \wp \in \mathcal{P}_{st}^{*}
\end{cases}
\end{align}
In other words, the RSP probability of a shortest path approaches the normalized likelihood of the path, which is also the contribution to the betweenness scores of the nodes along the path. 
Thus, the simple RSP betweenness converges to the shortest path likelihood betweenness defined in Section~\ref{sec:betweenness_centralities_sp}.

The same result holds for the RSP net betweenness. 
Intuitively, as the path distribution focuses more and more on the shortest paths, one of the two terms in the net flow in Equation (\ref{eq:randomized_shortest_path_net_betweenness01}) becomes zero, as the walker will only move in one direction along each edge for a given $s$-$t$-pair. 
Thus, the RSP net betweenness also approaches the shortest path likelihood betweenness, as $\beta \longrightarrow \infty$.

\subsection*{RSP betweenness centralities at the limit $\beta \to 0^{+}$}

In the high temperature limit, as $\beta \longrightarrow 0^{+}$, $\exp(-\beta \tilde{c}(\wp)) \longrightarrow 1$ for all $\wp$, and the RSP probabilities of Equation (\ref{eq:RSPprob}) converge to the unbiased random walk probabilities, determined by the reference transition probabilities, i.e.\
$
\widetilde{\mathrm{P}}_{st} \mathop{\longrightarrow}_{\beta \rightarrow 0^{+}} \Pref.
$
This means that the simple RSP betweenness converges to the expected number of visits to a node over all absorbing walks with respect to the unbiased random walk probabilities. 
As presented in Section~\ref{sec:betweenness_rw}, this measure is proportional to the stationary distribution, if the network is strongly connected and aperiodic, where the multiplicative factor is the sum of average hitting times between all $s$-$t$-pairs. 
Thus, as mentioned in Section~\ref{sec:betweenness_rw}, for undirected networks, the measure becomes proportional to the degree (or strength) centrality. 
In the same limit, as $\beta \longrightarrow 0^{+}$, the RSP net flow converges to the current flow betweenness centrality~\cite{Newman-05, Brandes-2005b}, as the edge flows $\overline{\eta}_{ij}(s,t)$ converge to the potential differences of adjacent nodes. 

\section{Experiments}
\label{sec:Exp}
The interpolation between common centrality measures already makes the simple RSP and RSP net betweenness centralities interesting. 
Furthermore, there are also cases in which the RSP betweenness centralities can be more relevant than their limit functions, the shortest path likelihood betweenness, the current flow betweenness and the stationary distribution. 
In this section these benefits will be illustrated, first with artificially generated networks, and later with two real networks of very different nature. 
The artificial examples show the behavior of the RSP betweenness measures in a network consisting of communities. 
The first real network is the street network of Lower and Midtown Manhattan, which serves as an example of an undirected network. 
The second is a subset of the directed Wikipedia hyperlink network.

All of the example cases presented here indicate benefits of using the RSP betweenness measures over the shortest path and random walk based betweenness measures. 
The benefits are clearer in the case of the simple RSP betweenness than the RSP net betweenness. 
The simple RSP betweenness is also more preferable in terms of interpretability and computational efficiency compared to the net betweenness. 
However, the decision of which approach to use depends on the actual application and its premises. 
We have also experimented with several other types of graphs, including, for instance, examples presented by Brandes and Hildenbrand~\cite{brandes2014smallest}, which have been designed to differentiate centrality measures. 
However, for these, and other simple cases, there is no clear difference in considering RSPs, but rather the ranking of nodes with intermediate values of $\beta$ are in essence the same as with the limit values of $\beta$.

\subsection{Overall betweenness of an in-between community}

One possible use for the RSP betweenness measures is the detection of groups of nodes that are central in a network. 
Consider a network consisting of three disjoint communities, A, B and C which are highly intraconnected but loosely interconnected. 
In addition, community B is connected to communities A and C, which, however, do not share any edges with each other. 
In other words, community B is in between communities A and C and all paths between nodes of communities A and C have to go through community B. 
Such an organization is possible, for instance, in a hierarchical social network, where community B could represent a directoral board of a company. 
If, moreover, the graph is in general sufficiently sparse, then the shortest paths between communities A and C will run through only a few of the nodes of community B. 
Thus, betweenness measures based on shortest paths will only highlight those nodes of community B, whereas the other nodes of community B will get no contribution from the connections between communities A and C. 
For some applications, however, the nodes of community B should, in general, be considered more in-between than the nodes of communities A and C. 
Defining betweenness based on random walks, and especially RSP's can help in this matter, as will be demonstrated next.
 
\begin{figure}[t]
\begin{center}

\subfigure[Shortest path]{
\includegraphics[scale=.5]{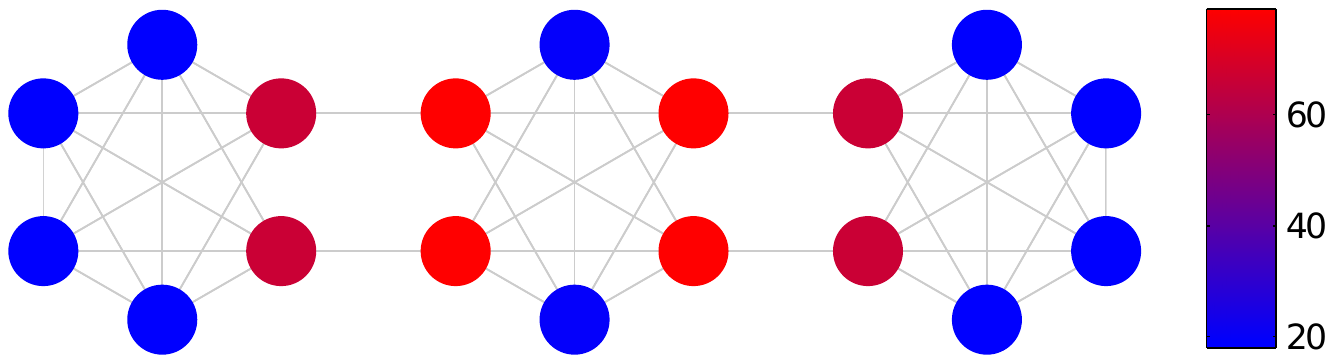}
\label{fig:3comms1}
}
\subfigure[Simple RSP, $\beta=0.01$]{
\includegraphics[scale=.5]{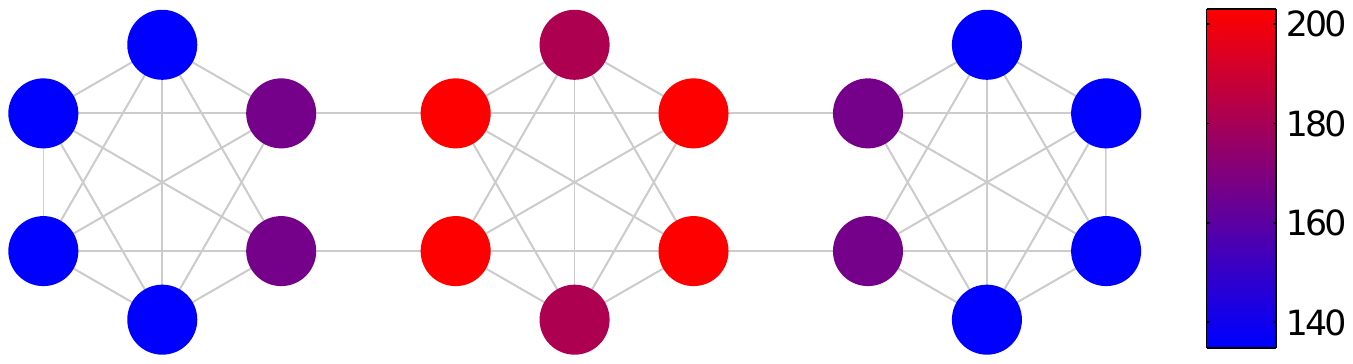}
\label{fig:3comms2}
}
\subfigure[Degree]{
\includegraphics[scale=.5]{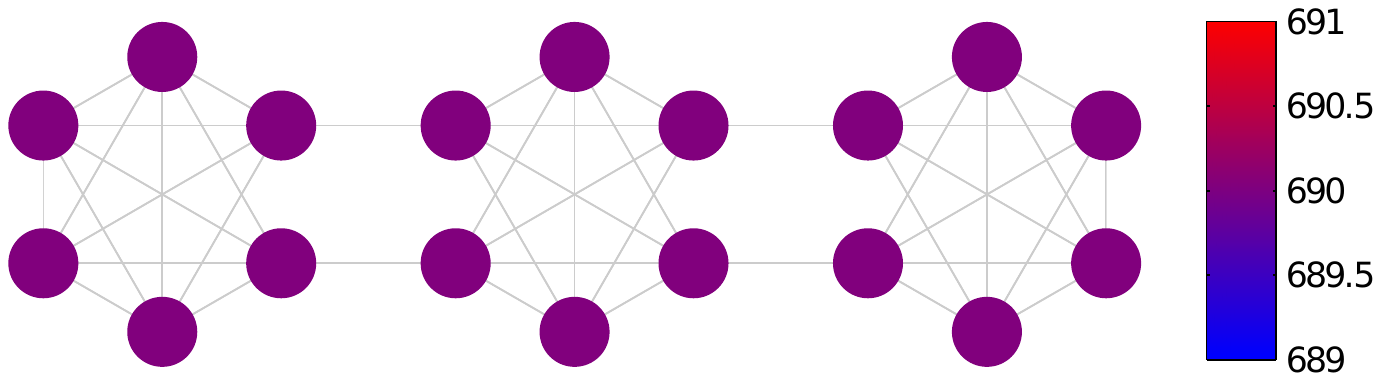}
\label{fig:3comms3}
}
\caption{A 5-regular graph with three communities, and the heat plots of its betweenness values with the shortest path likelihood betweenness (a), the simple RSP betweenness (b) and the degree centrality. 
Blue indicates low, and red high betweenneess values.}
\label{fig:threeComms_illustration}
\end{center}
\end{figure}

We first consider a simple example of a regular graph with communities organized in the order described above. 
The example is depicted in Figure~\ref{fig:threeComms_illustration} which shows a 5-regular graph containing three communities of 6 nodes, where one of the communities is connected to the other two. 
The graph has been constructed by considering three cliques of 6 nodes, and by removing and adding appropriate links to obtain the desired structure. 
Figure~\ref{fig:threeComms_illustration} contains the heat plots of the betweenness values of the nodes with the simple RSP betweenness (Figure~\ref{fig:threeComms_illustration}(b)) and its limit functions, i.e.\ the shortest path likelihood betweenness (Figure~\ref{fig:threeComms_illustration}(a)), which in this example equals the standard shortest path betweenness, and the stationary distribution multiplied by the sum of average hitting times (see section ``RSP betweenness centralities at the limit $\beta \longrightarrow 0^+$''), which -- as the network is undirected -- corresponds to the degree centrality (Figure~\ref{fig:threeComms_illustration}(c)) up to a scaling factor.

The heat plots show that the simple RSP betweenness highlights the nodes of the central community more than its limit functions. 
The low temperature limit function, i.e.\ the shortest path likelihood betweenness highlights the nodes connecting the different communities, but the betweenness scores of the two other nodes in the central community are of the same magnitude as the scores of the other nodes in the peripheral communities. 
In the high temperature limit the simple RSP betweenness converges to the degree centrality, which is constant for all nodes, as the graph is regular. 
Although the heat plots show the actual betweenness scores, and not the rankings of the nodes according to them, the rankings also comply with the above findings.

Using the RSP net betweenness (for which the results are not illustrated here), however, brings no benefit in this example, when compared to its limit functions. 
Namely, the current flow betweenness ranks the nodes of the central community a bit higher than the shortest path likelihood betweenness, but the RSP net betweenness does not increase those ranks with any intermediate values of $\beta$. 
On the other hand, the current flow betweenness values of the central community nodes are relatively much lower than the simple RSP betweenness values of Figure~\ref{fig:threeComms_illustration}(b). 
However, the RSP net betweenness, can be beneficial in a similar setting, but with a bit of more complexity, which will be shown next. 

For a more complex example, we generate random networks with the LFR algorithm of Lancichinetti et al.\ \cite{lancichinetti2008benchmark} designed to construct scale-free networks with a community structure. 
This experiment confirms further the usefulness of the simple RSP, as well as the RSP net betweenness measure. 
We generated graphs consisting of three communities, A, B and C, and then simply removed the edges between two of the communities A and C. 
The size of the communities was set to 120 nodes, resulting in networks with 360 nodes, the average degree of the network was set to 10, the maximum degree to 120 and the power-law exponent of the degree distribution to $-2$. 
We tested three different values of the mixing coefficient $\mu = \{0.01, 0.05,0.1\}$, which essentially determines the probability of having an edge between two communities after the degrees of nodes have been fixed. 
For each generated network, we computed the shortest path likelihood betweenness, the degree centrality, the current flow betweenness as well as the simple RSP and RSP net betweenness scores with several different values of the parameter $\beta$. 
We then rank the nodes according to each list of betweenness scores (with rank 1 naturally meaning the node with the highest score) and compute the average rank of the nodes in the central community B. 
We repeat the graph generation and the computation of average ranks 200 times and report the mean average rank of the nodes in the in-between community over these 200 runs.

\begin{figure*}[]
\begin{center}

\subfigure[]{
	\includegraphics[scale=.62, trim=60 350 105 250, clip=true]{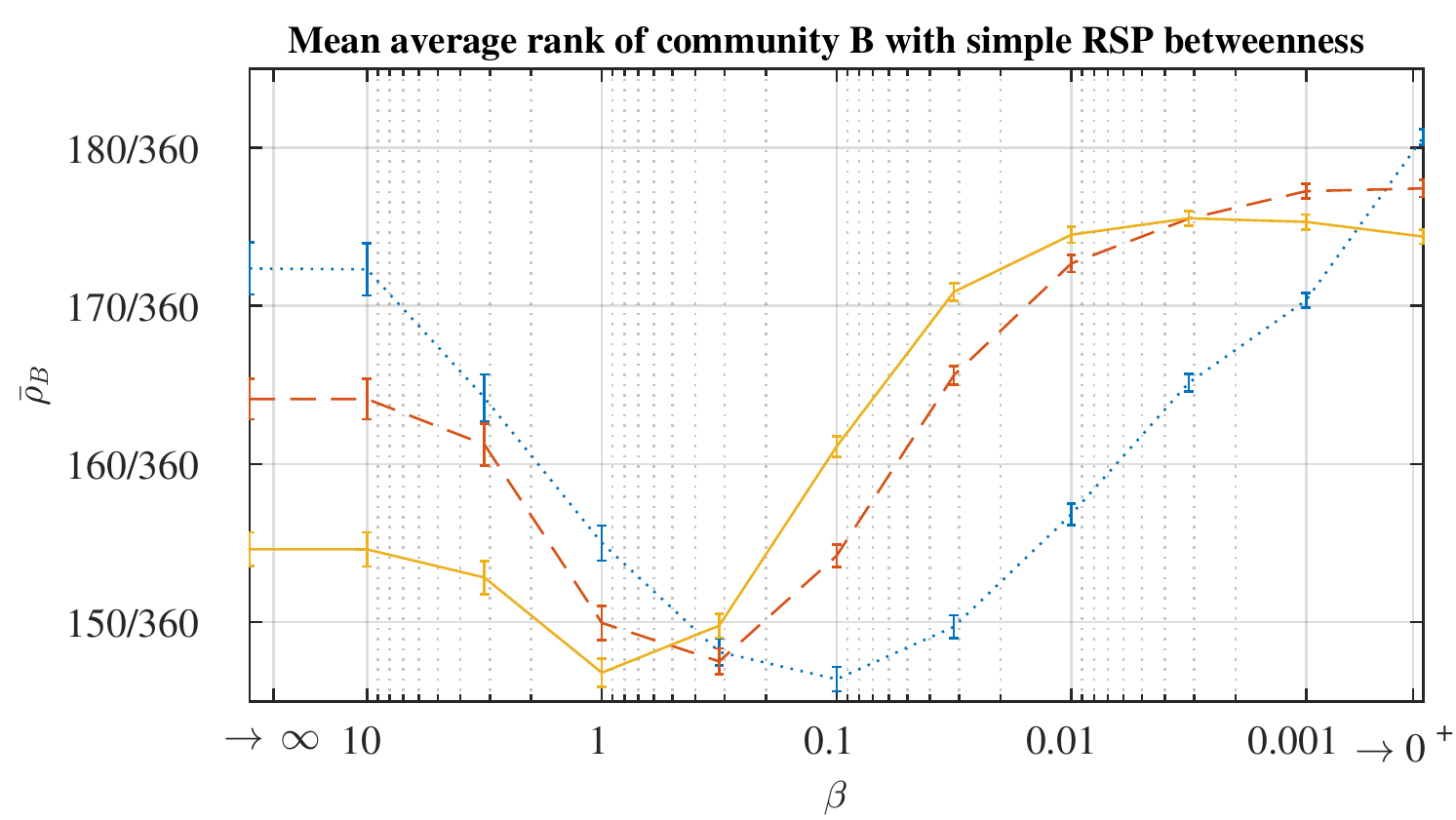}
	\label{fig:threeComms_a}
	}
\subfigure[]{
	\includegraphics[scale=.62, trim=125 350 105 250, clip=true]{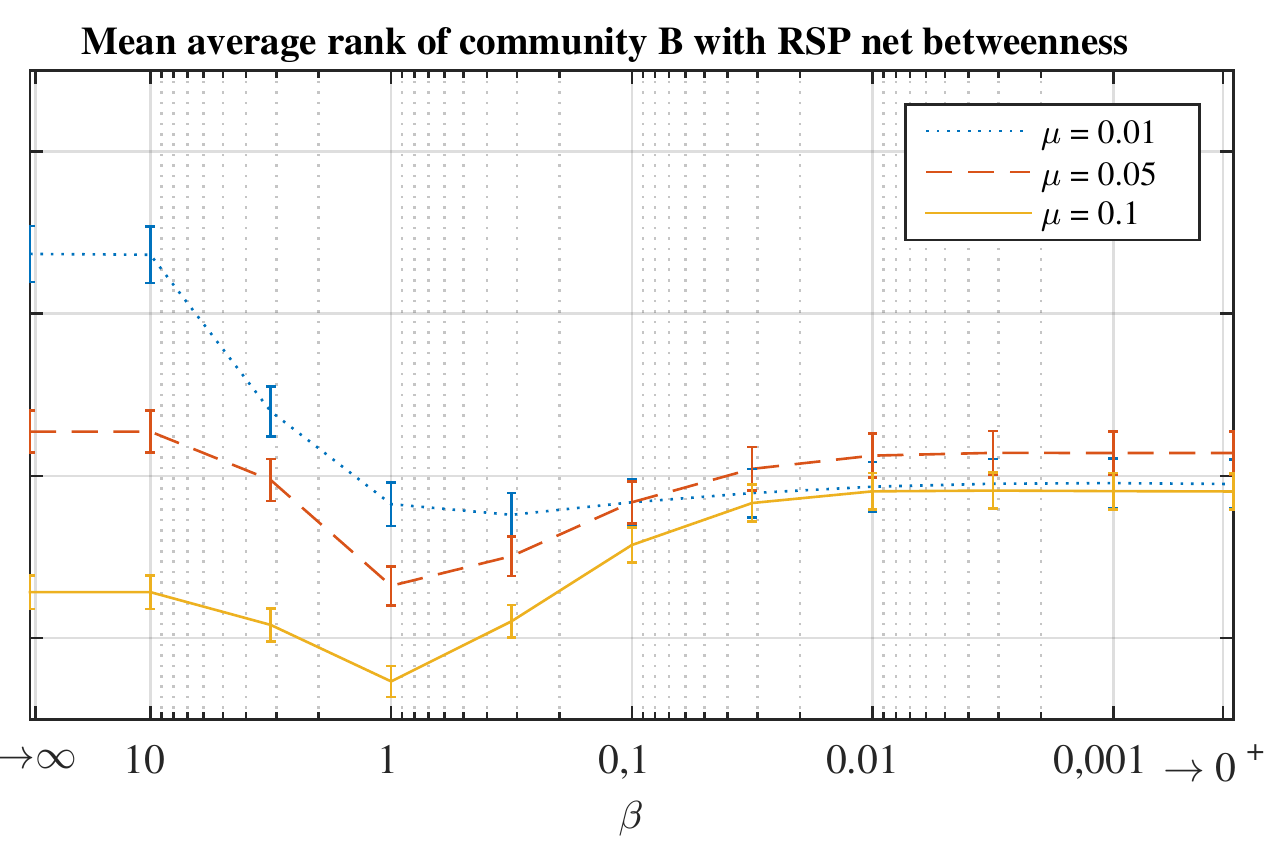}
	\label{fig:threeComms_b}
	}
	\caption{The mean average rank $\bar{\rho}_B$ of the nodes of community B which lies in between two other communities, A and C, based on the nodes' RSP betweenness (a) and RSP net betweenness values (b) over 200 networks of 360 nodes generated using the LFR algorithm, as described in the body text (with low rank meaning a high betweenness score). 
The results are plotted for varying values of $\beta$ and for three values of the mixing parameter $\mu$ with error bars indicating the standard error of the mean over the 200 runs. 
In both plots, the values at the left end of the curves (as $\beta \rightarrow \infty$) show the results with the shortest path likelihood betweenness and at the right end (as $\beta \rightarrow 0^+$) the results with the degree centrality in (a), and the current flow betweenness in (b).}
\label{fig:threeComms}
\end{center}
\end{figure*}

The results are plotted in Figure~\ref{fig:threeComms}, which shows that in most cases both the simple RSP and the RSP net betweenness, with some intermediate values of $\beta$, rank the nodes of the in-between community more central than their limiting functions. 
The plots show the mean average rank, as well as the standard error of the mean, of the nodes of the central community B with different values of the inverse temperature $\beta$. 
The values at the extremes of the plots correspond to the results obtained with the limiting functions, with the left end corresponding to the low-temperature (high $\beta$) case, i.e.\ the shortest path likelihood betweenness in both plots, and the right end to the high-temperature (low $\beta$) case, i.e.\ the degree centrality in Figure \ref{fig:threeComms_a}, and the current flow betweenness Figure \ref{fig:threeComms_b}. 
It is evident from the figures that with some intermediate value of $\beta$ the RSP betweenness measures in this setting often manage to rank the nodes of the central group B higher than the limit functions by taking into account other connections besides only the shortest ones. 
Note that this does not mean that the RSP betweenness rankings are lower than the rankings with the limit functions on each individual network, but that the result holds on average over the 200 generated networks. 
The fluctuations of the rankings are indicated by the error bars showing the standard error of the mean over the 200 networks. 

\subsection{Manhattan street network}

\newcommand{\ManhSize}{.17}
\newcommand{\ManhScale}{.17}
\newcommand{\ManhTrim}{0 0 0 0}

\begin{figure*}[t]

\begin{center}
\begin{minipage}[t][][c]{0.23\linewidth}

\vspace*{1.6cm}
\subfigure[Shortest path]{
\label{fig:RSP_manhattan_SP}
\includegraphics[clip=true, trim=40 100 20 60, scale=.21]{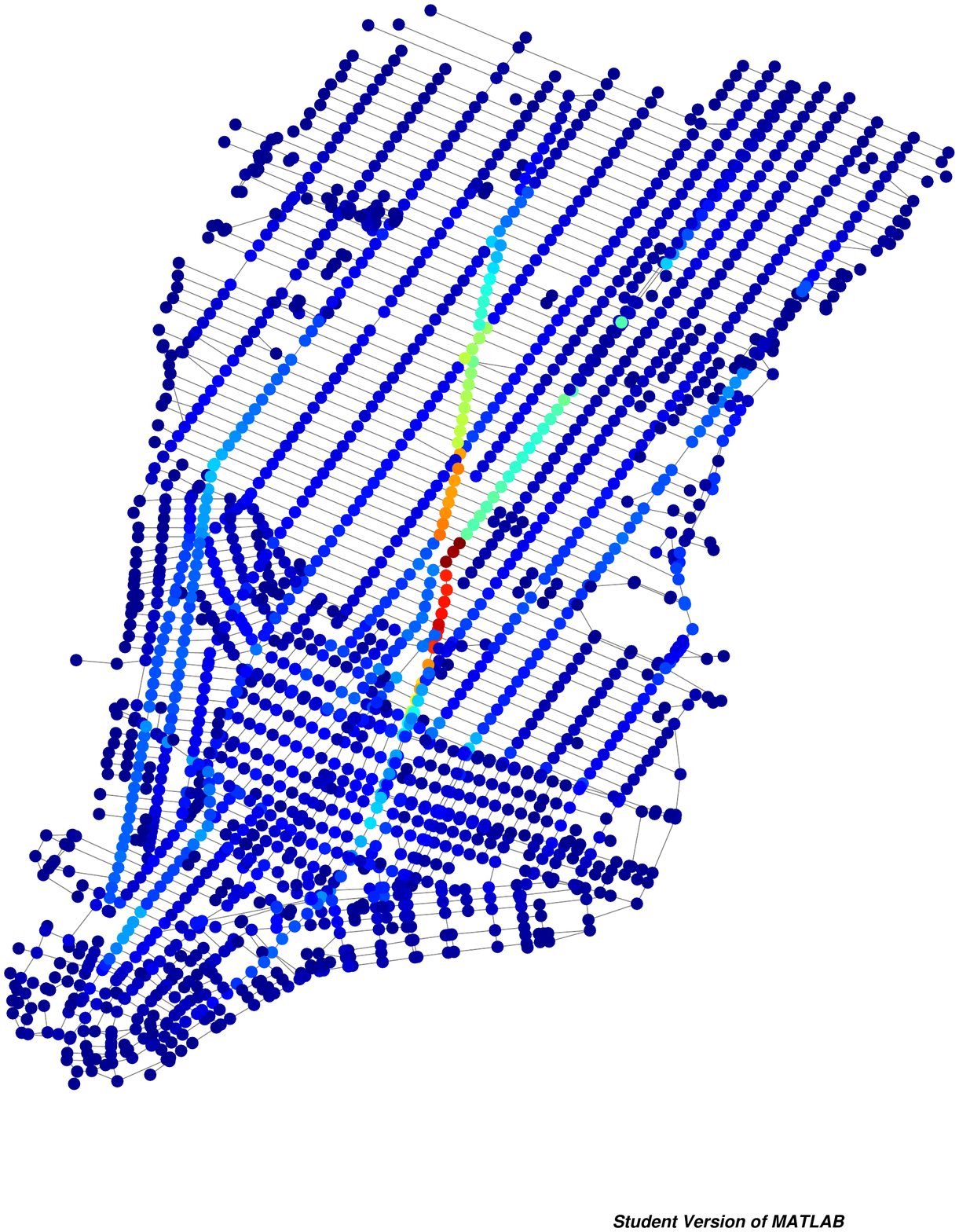}}

\end{minipage}
\begin{minipage}[t][][c]{\ManhSize\linewidth}

\subfigure[RSP, $\beta=10^{-2}$]{
\label{fig:RSP_manhattan_RSP1}
\includegraphics[clip=true, trim=20 100 20 60, scale=\ManhScale]{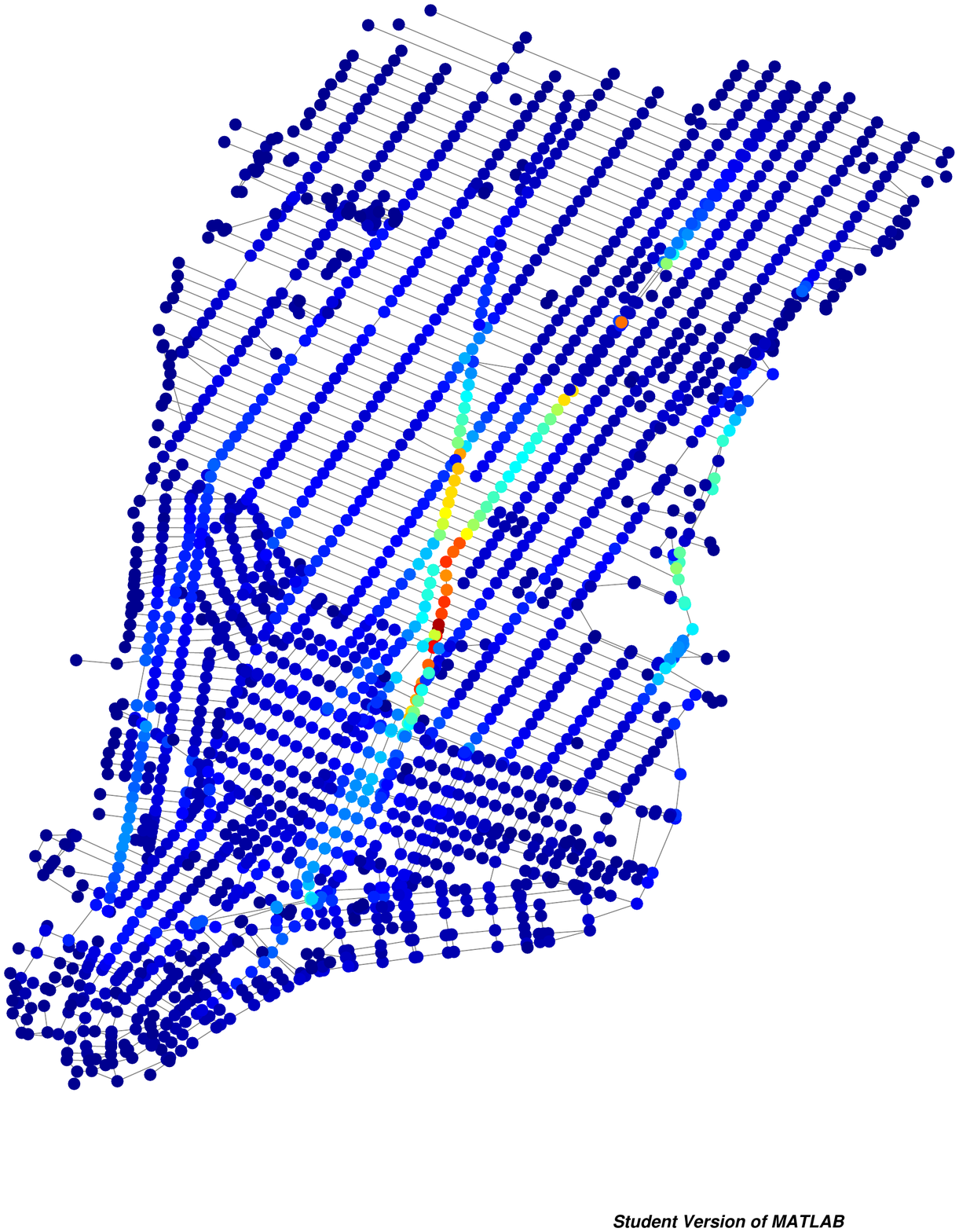}}

\subfigure[RSP net, $\beta=10^{-2}$]{
\label{fig:RSP_manhattan_RSPN1}
\includegraphics[clip=true, trim=20 100 20 60, scale=\ManhScale]{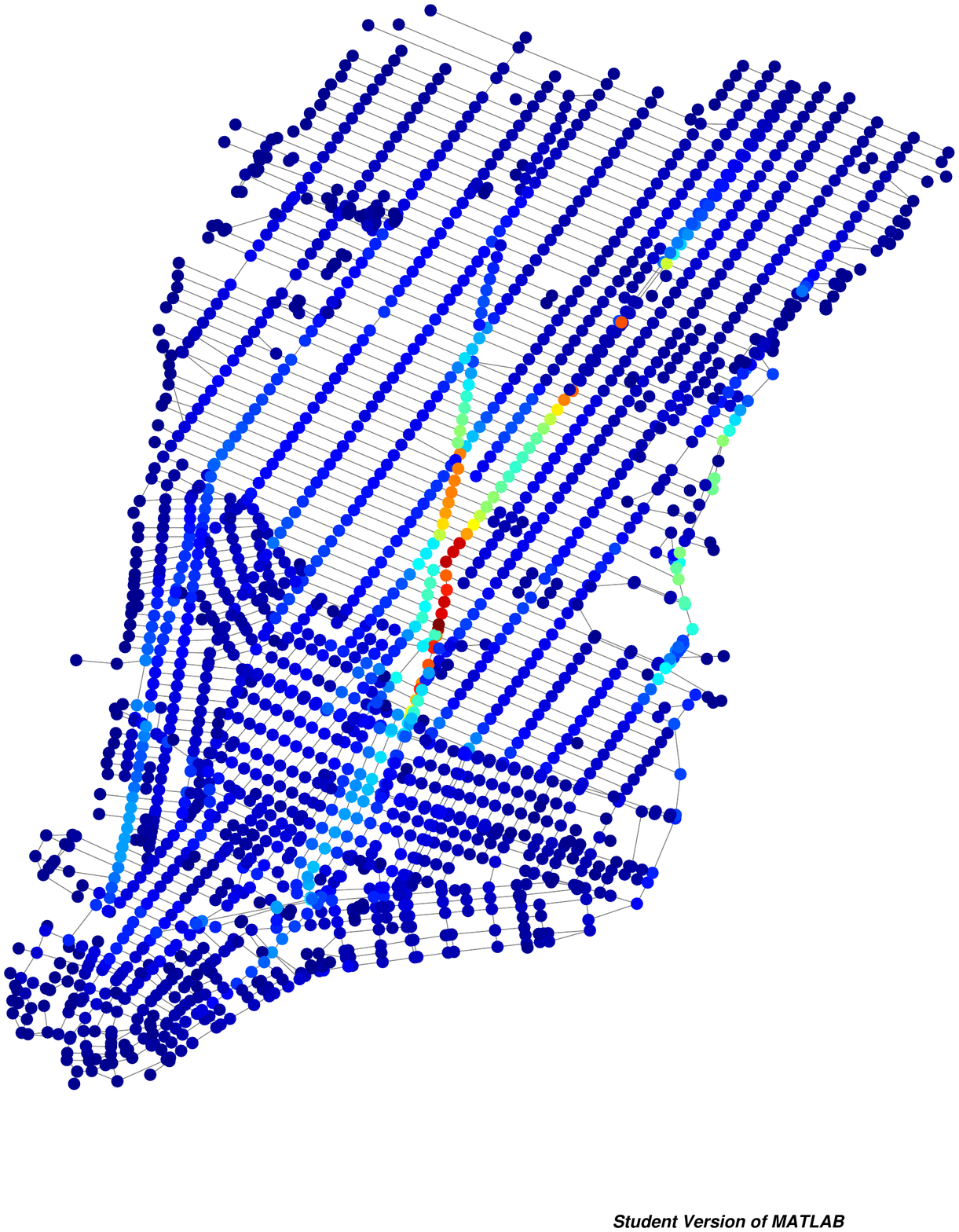}}

\end{minipage}
\begin{minipage}[t][][c]{\ManhSize\linewidth}

\subfigure[$\beta=10^{-3}$]{
\label{fig:RSP_manhattan_RSP2}
\includegraphics[clip=true, trim=20 100 20 60, scale=\ManhScale]{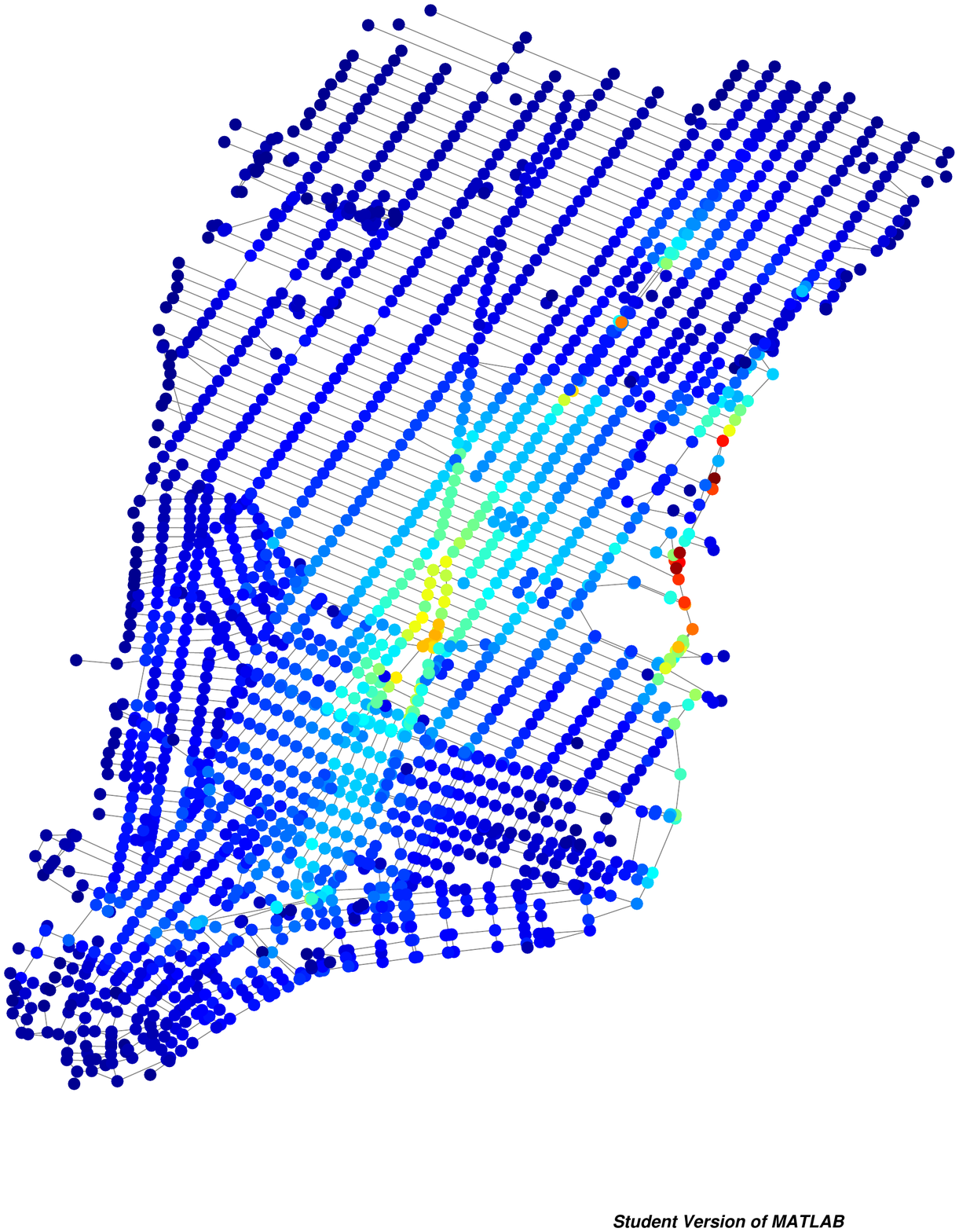}}

\subfigure[$\beta=10^{-3}$]{
\label{fig:RSP_manhattan_RSPN2}
\includegraphics[clip=true, trim=20 100 20 60, scale=\ManhScale]{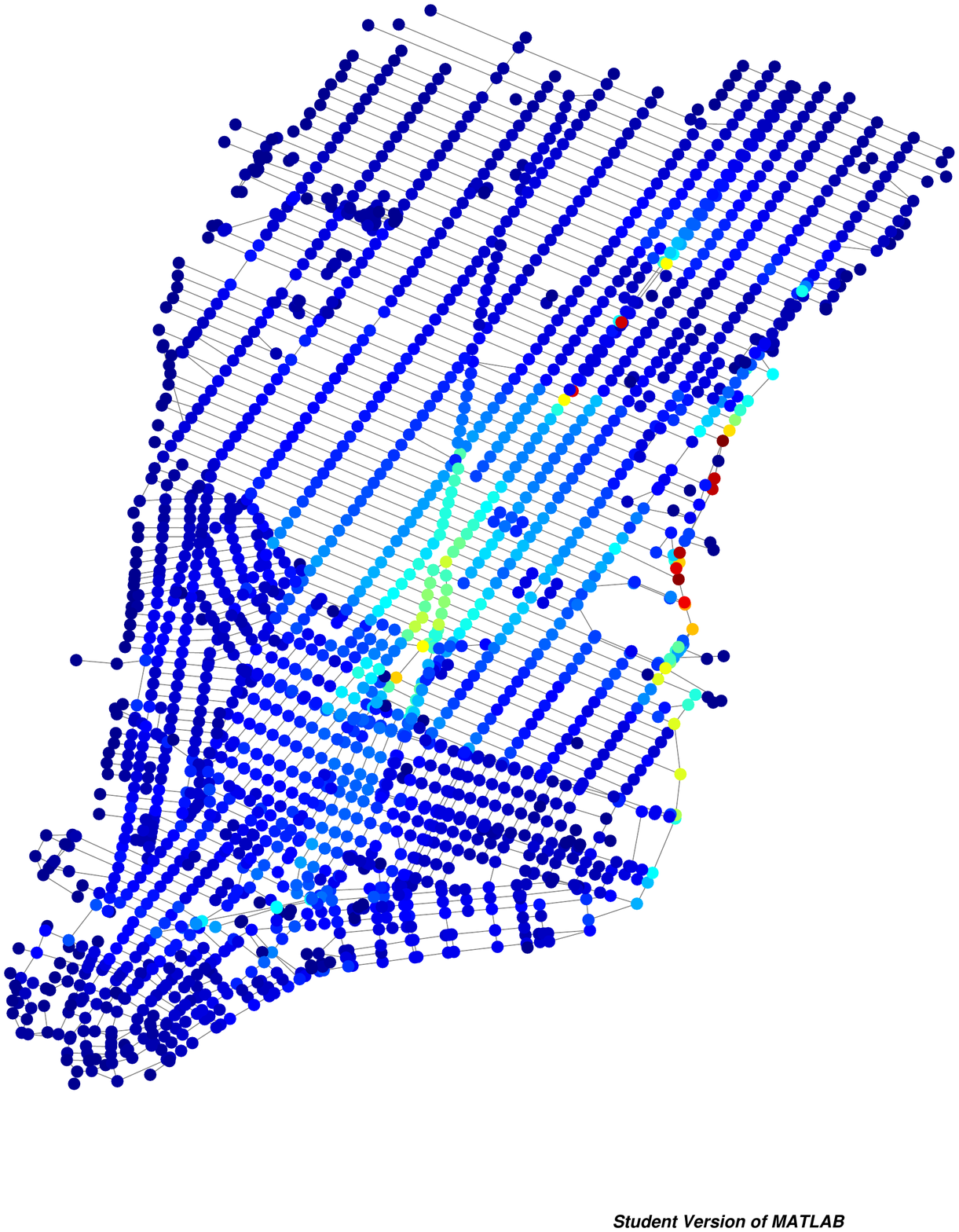}}

\end{minipage}
\begin{minipage}[t][][c]{\ManhSize\linewidth}

\subfigure[$\beta=10^{-4}$]{
\label{fig:RSP_manhattan_RSP3}
\includegraphics[clip=true, trim=20 100 20 60, scale=\ManhScale]{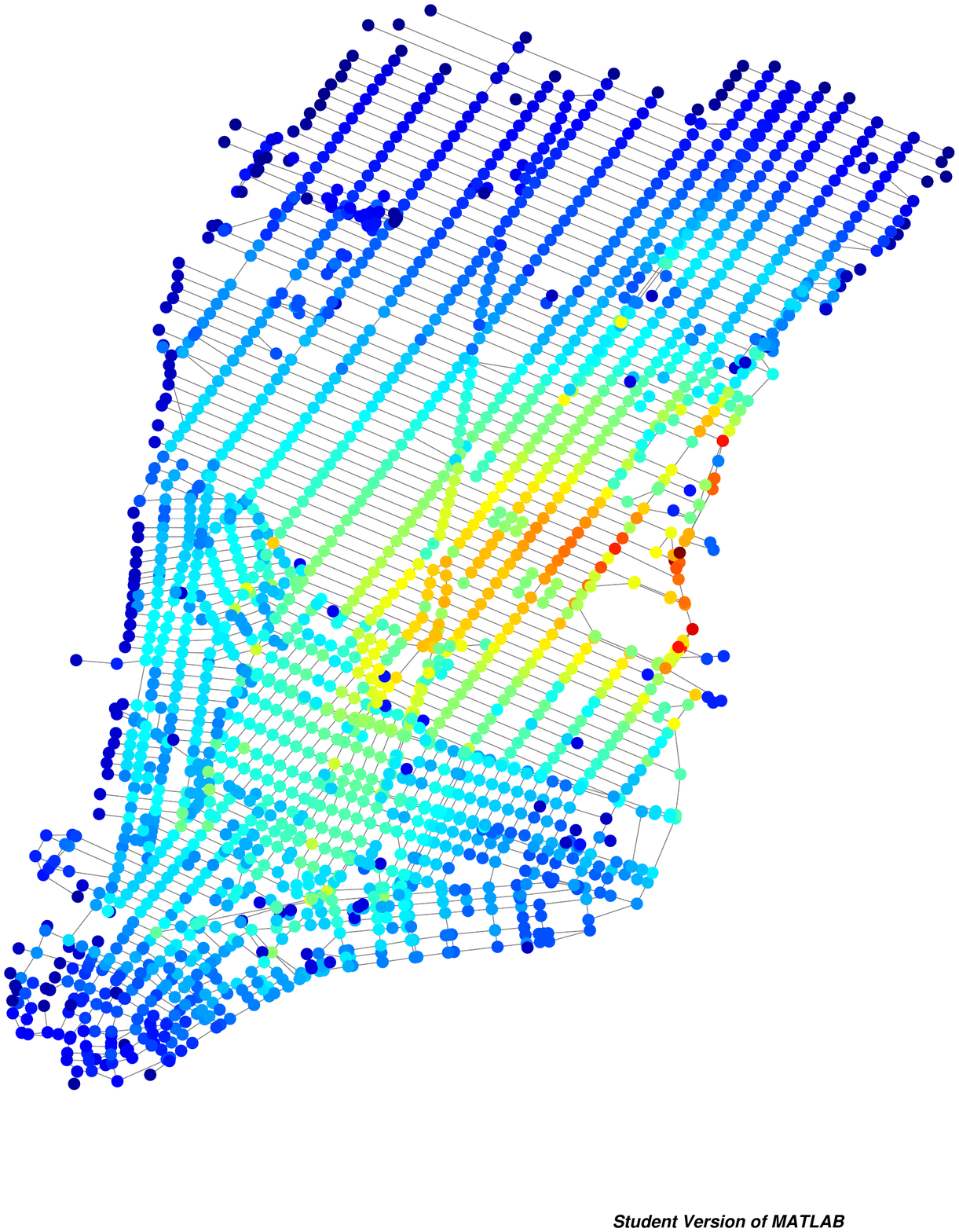}}

\subfigure[$\beta=10^{-4}$]{
\label{fig:RSP_manhattan_RSPN3}
\includegraphics[clip=true, trim=20 100 20 60, scale=\ManhScale]{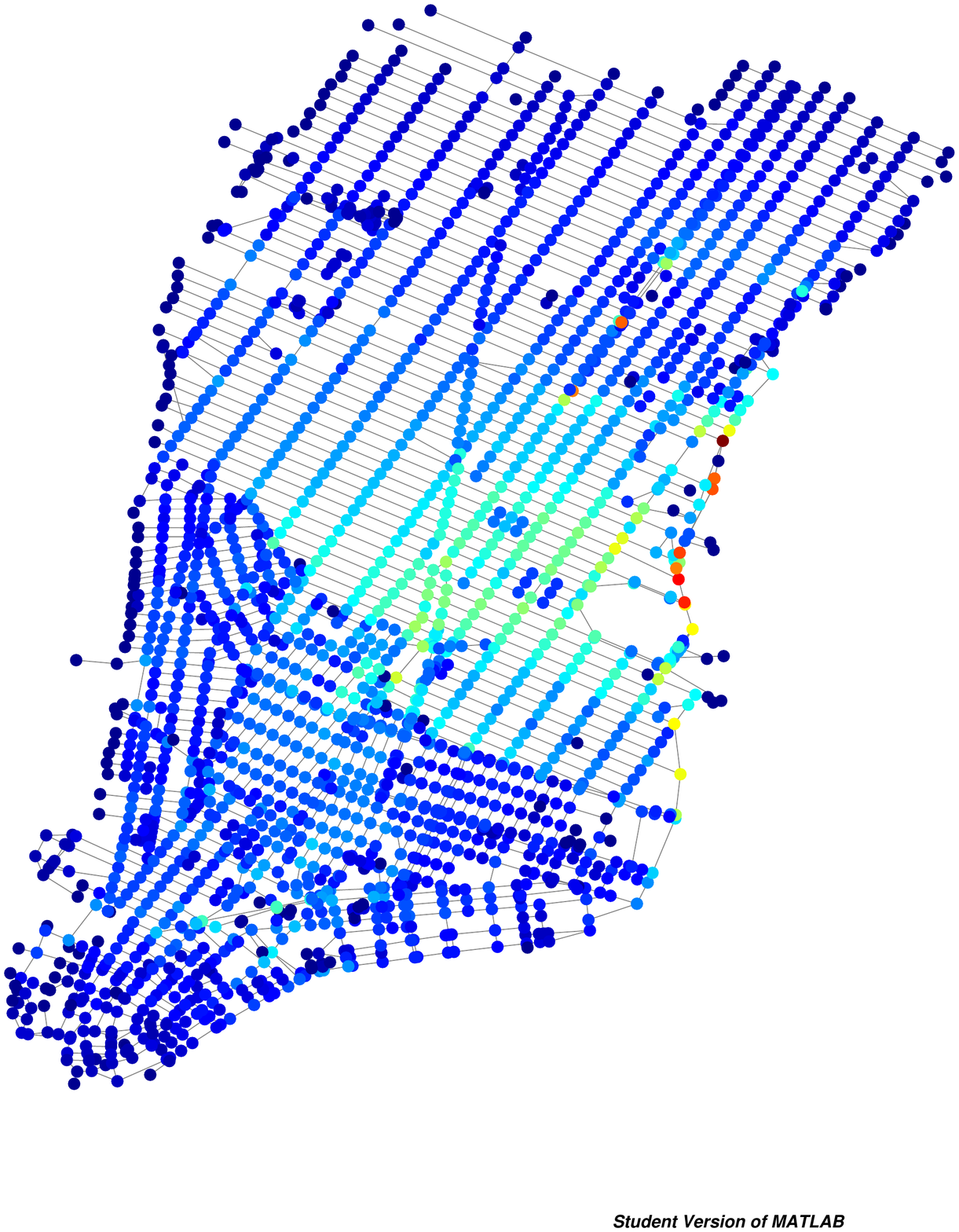}}

\end{minipage}
\begin{minipage}[t][][c]{\ManhSize\linewidth}

\subfigure[Degree]{
\label{fig:RSP_manhattan_stat}
\includegraphics[clip=true, trim=20 100 20 60, scale=\ManhScale]{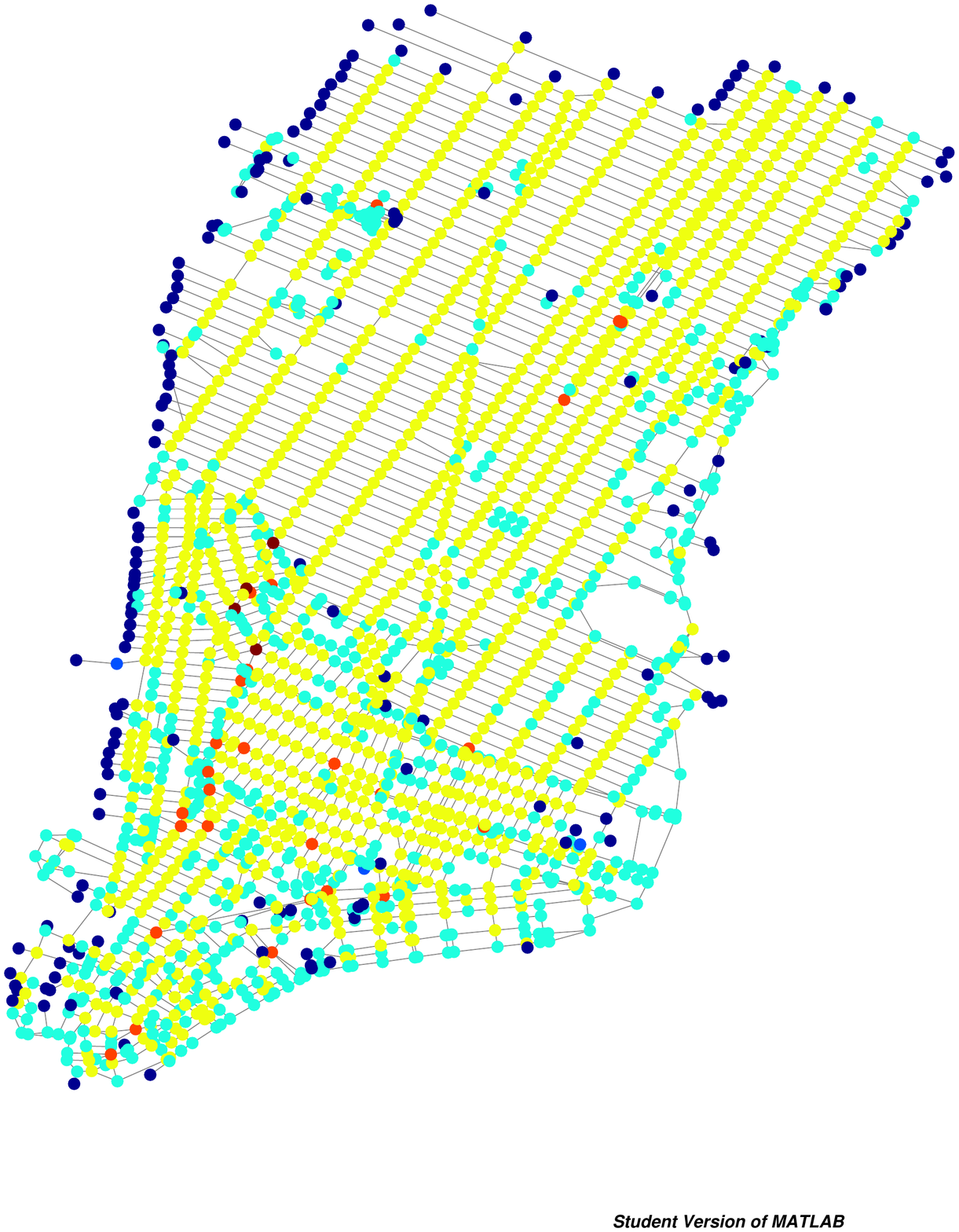}}

\subfigure[CFB]{
\label{fig:RSP_manhattan_CF}
\includegraphics[clip=true, trim=20 100 20 60, scale=\ManhScale]{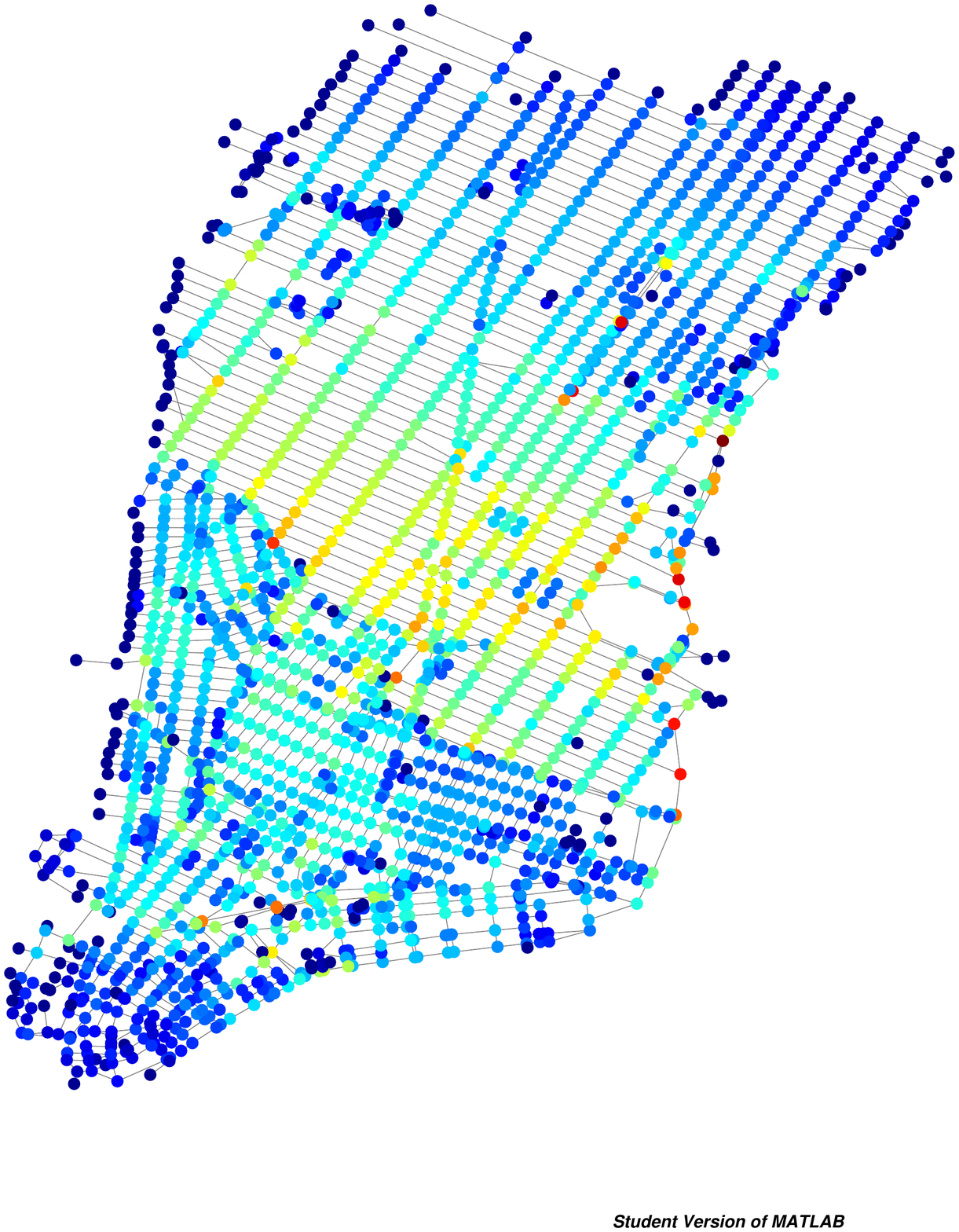}}

\end{minipage}
\end{center}
\caption{The interpolation between the shortest path and random walk betweenness measures with the simple RSP and the RSP net betwenness measures on the Midtown and Lower Manhattan street network. 
The network has been extracted from \href{http://www.openstreetmap.org/}{OpenStreetMap}\cite{OSM} (for the data and code, see Materials).}
\label{fig:RSP_manhattan}
\end{figure*}

One promising application area for RSP's are routing or path planning problems. 
RSP's allow the modeling of routing in situations that include an element of randomness, such as navigation of people or animals in an environment. 
On the other hand, by definition, RSP's can be used for planning paths in an optimal way while keeping the predictability of the path at a desired level. 
RSP's could also be used for avoiding congestion problems in transportation and traffic networks. 

We illustrate the use of RSP's for routing in a network by analyzing the street network of Midtown and Lower Manhattan. 
We have extracted this network from \href{http://www.openstreetmap.org/}{OpenStreetMap}\cite{OSM}. 
The nodes in the network correspond to intersections and the edges are the street segments between the intersections. 
We treat the network as undirected, and analyze the network using both the simple RSP and the RSP net betweenness. 

The length of each street segment is assigned as the cost of the corresponding edge. 
Accordingly, the overall cost of a path is its overall length. 
However, we define here the reference transition probabilities of the random walk according to the degree of each node, $p_{ij}^{\mathrm{ref}} = 1/d_i$, i.e.\, only according to the number of edges connected to the node and independent of the edge costs. 
This seems like a reasonable choice for moving in a street network, if we consider that the decision of direction of the random walker in an intersection is not affected by the lengths of the street segments. 
Remember that this means that the shortest path likelihood betweenness on the graph is based on the edge costs, whereas the random walk based betweenness measures do not depend on the costs, but only the degrees of nodes.

The heat plots of the simple RSP and the RSP net betweenness measures and their limit functions on the Manhattan street network are depicted in Figure~\ref{fig:RSP_manhattan}. 
At low temperatures (large $\beta$), both RSP based betweenness measures converge to the shortest path likelihood betweenness, shown in Figure~\ref{fig:RSP_manhattan_SP}. 
Figures \ref{fig:RSP_manhattan_RSP1}, \ref{fig:RSP_manhattan_RSP2} and \ref{fig:RSP_manhattan_RSP3} show the betweenness values obtained with the simple RSP betweenness, and Figures \ref{fig:RSP_manhattan_RSPN1}, \ref{fig:RSP_manhattan_RSPN2} and \ref{fig:RSP_manhattan_RSPN3} the values obtained with the RSP net betweenness with different values of $\beta$. 
As the network is undirected and connected, and because we use reference probabilities based only on degrees, instead of costs, the limit function of the simple RSP betweenness, when $\beta \longrightarrow 0^{+}$, is equal to the degree centrality multiplied by a constant, shown in Figure~\ref{fig:RSP_manhattan_stat}. 
Finally, the limit function of the RSP net betweenness is the current flow betweenness, which is presented in Figure~\ref{fig:RSP_manhattan_CF}.

This example shows one strength of both of the RSP betweenness measures. 
It is evident from the plot that the shortest path betweenness is focused on Broadway, which functions as a diagonal shortcut in many routes on the grid-like Midtown. 
However, when $\beta$ is decreased, both RSP based measures rank highest the intersections along the FDR Drive on the eastern shore. 
This is mainly caused by the sparsity of streets on the east shore close to the residential areas of Stuyvesant Town and Peter Cooper Village. 
As a result, the FDR Drive is a vital connection between the upper and lower eastern parts of the map. 
This aspect is not clear from the shortest path likelihood betweenness, but becomes apparent by computing the RSP-based betweenness values. 
The current flow betweenness also ranks high the intersections of FDR Drive, but as a drawback the importance of Broadway is not as apparent as it perhaps deserves.

Thus, it seems that the RSP based betweenness measures, by assuming suboptimal navigation between points in the network, can highlight bottlenecks such as the FDR Drive on the Manhattan network better than the deterministic shortest path betweenness or the unbiased current flow betweenness. 
Comparing between the two RSP betweenness, it is hard to find any major differences. 
However, in a street network, when considering the movement of people or vehicles, the simple RSP betweenness has a more sensible physical interpretation than the net betweenness. 
Moreover, again, the computation of the simple RSP betweenness is much more efficient and its interpretation clearer than the net betweenness, which make it a more preferable candidate. 
On the other hand, there may also appear applications, for which the net flow interpretation is more relevant, in which case the net betweenness should be used.

\subsection{A subnetwork of Wikipedia}

\newcommand{\WikiWidth}{0.3\linewidth}

\begin{figure*}[t]
\begin{center}
\subfigure[Shortest path]{
\label{fig:Wiki_SP}
\includegraphics[width=\WikiWidth]{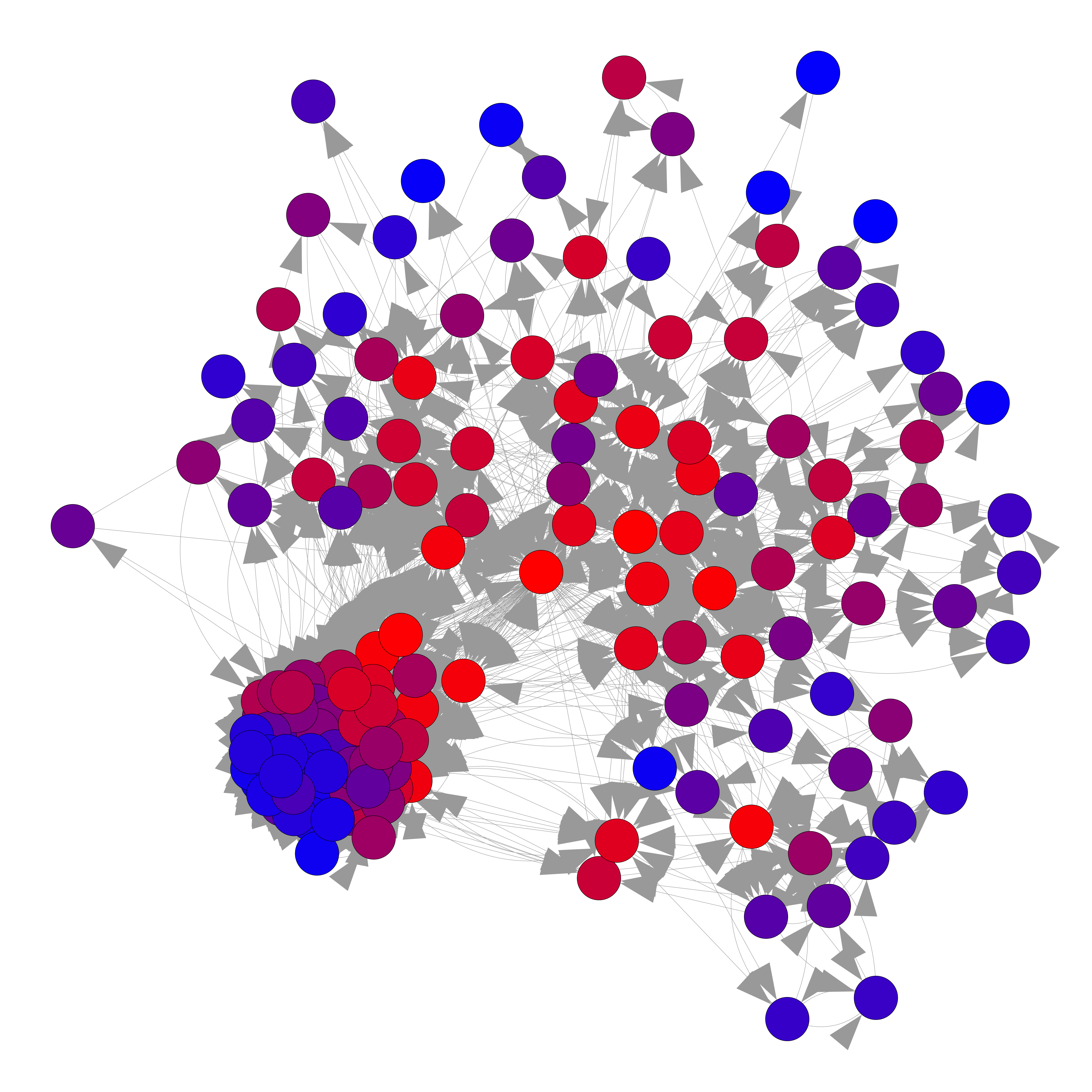}}
\subfigure[RSP, $\beta=10^{-1}$]{
\label{fig:Wiki_RSP}
\includegraphics[width=\WikiWidth]{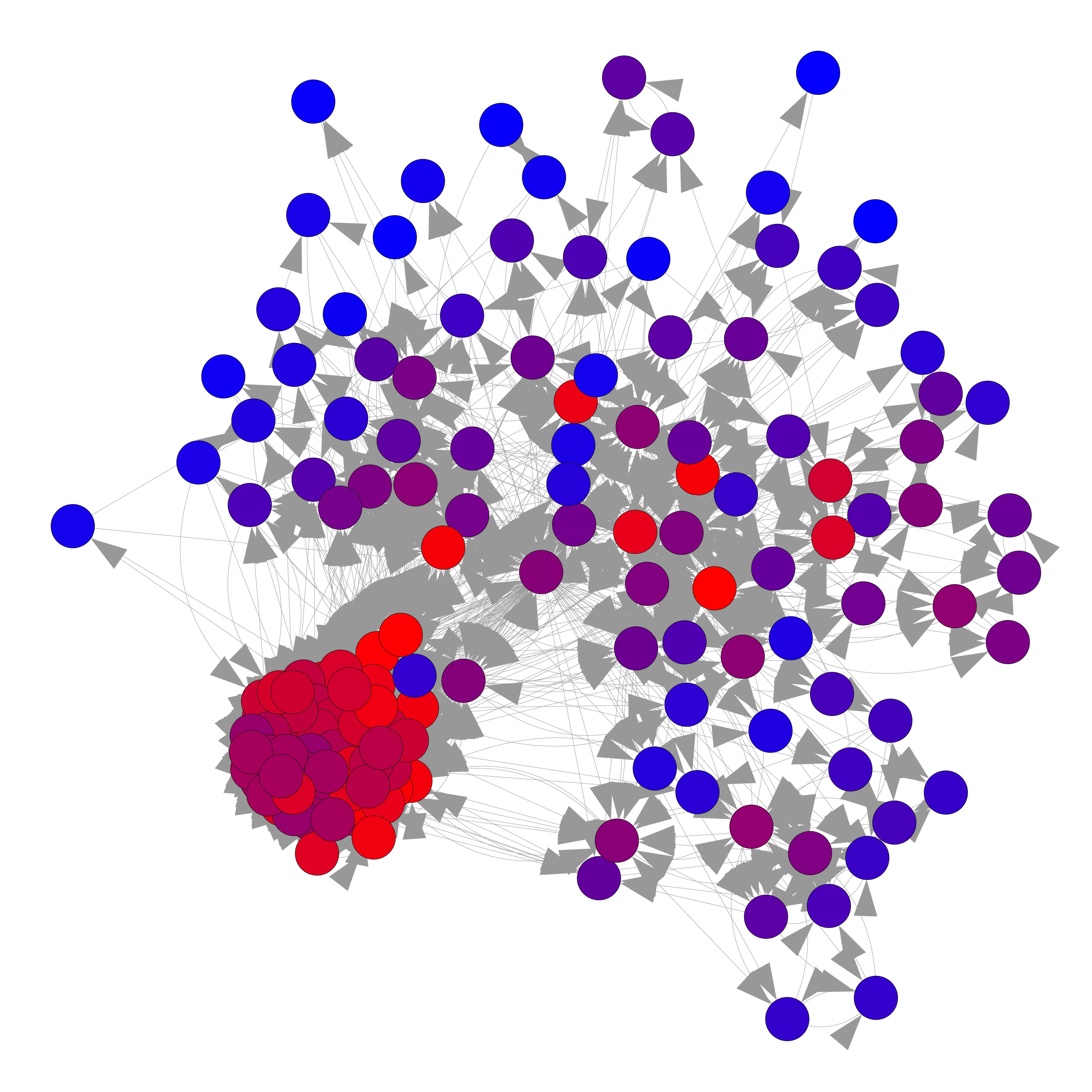}}
\subfigure[Stationary distribution]{
\label{fig:Wiki_Stat}
\includegraphics[width=\WikiWidth]{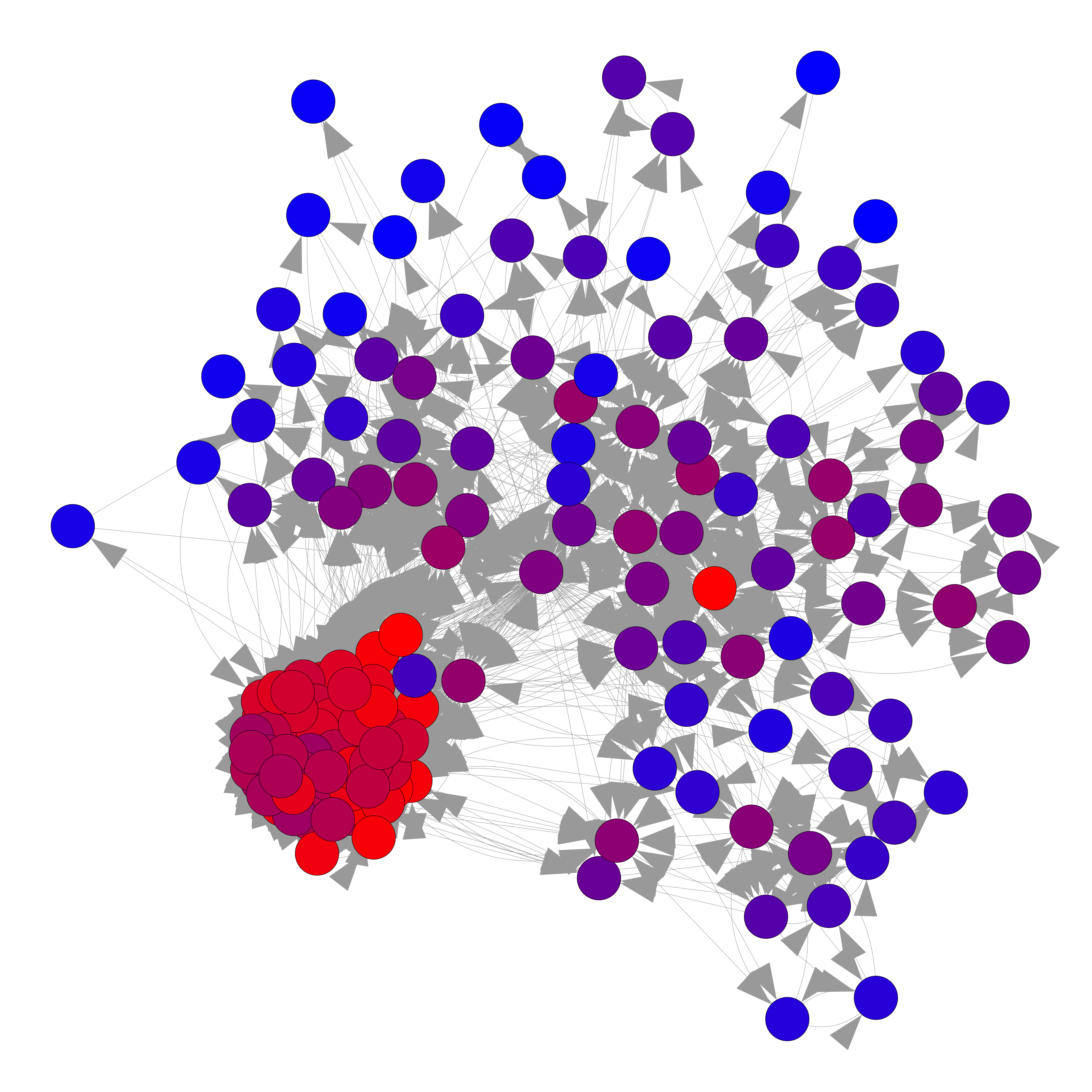}}

\footnotesize{
\begin{tabular}{p{1.8cm}p{5.3cm}p{5.0cm}p{4.8cm}}

& 1.\ Network science           & 1.\ Graph theory              & 1.\ Graph theory               \\
& 2.\ Network theory            & 2.\ Social network            & 2.\ Social network             \\
& 3.\ Social network            & 3.\ Social network analysis   & 3.\ Social networking service  \\
& 4.\ Graph theory              & 4.\ Mathematics               & 4.\ Myspace                    \\
& 5.\ Social network analysis   & 5.\ Sociology                 & 5.\ Social network analysis    \\
& 6.\ PageRank                  & 6.\ Network theory            & 6.\ Pinterest                   \\
& 7.\ Small-world network       & 7.\ Computer science          & 7.\ Orkut\\
& 8.\ Sociology                 & 8.\ Social networking service & 8.\ Small world experiment\\
& 9.\ Small world experiment    & 9.\ PageRank                  & 9.\ Tumblr\\
&10.\ Orkut                     &10.\ Network science           &10.\ Social network analysis software
\end{tabular}
}
\end{center}

\caption{The interpolation between the shortest path likelihood betweenness the stationary distribution with the simple RSP betwenness on the subnetwork of Wikipedia. 
The color of a node in the plots indicates its rank w.r.t.\ the betwenness measure; red and blue indicate high and low rank (i.e.\ high and low betweenness values), respectively. 
The dense cluster of nodes in the lower left corner consists mostly of nodes related to social networks. 
Below the plots, the 10 highest-ranked nodes are listed.}
\label{fig:Wiki}
\end{figure*}

Here we illustrate the behavior of the simple RSP betweenness on a directed real network. 
The network is a subnetwork of the hyperlink network of Wikipedia. 
It consists of the Wikipedia page on Network Science and the pages that contain a hyperlink to it, or are linked to from it. 
We only consider the largest strongly connected component of this network which contains 151 nodes. 
We only report the results obtained with the simple RSP betweenness, and not the RSP net betweenness, because the net flow interpretation is not particularly suitable when studying the World Wide Web, and in this example it does not provide any interesting results.

Figure \ref{fig:Wiki} shows the change in the rankings of the nodes of the Wikipedia subnetwork with the simple RSP betweenness centrality (Figure~\ref{fig:Wiki_RSP}) as well as with its limiting functions, the shortest path likelihood betweenness (Figure~\ref{fig:Wiki_SP}) and the stationary distribution (Figure~\ref{fig:Wiki_Stat}). 
In addition to the heat plots, there are lists of the top ten nodes according to each betweenness centrality. 
The plots directly illustrate the general structure of the network, with a division to a tightly intra-connected cluster, appearing on the lower left corner of the plots, and a more sparsely connected peripheral part. 
The dense cluster comprises of nodes related mostly to social networks, whereas the other nodes correspond to more general concepts.

The shortest path likelihood betweenness ranks quite high many nodes from both of the two groups of nodes, i.e.\ general nodes such as the seed node 'Network science', whereas the stationary distribution focuses mostly on the dense cluster of nodes related to social networks, highlighting especially particular social networking websites, such as 'Myspace' and 'Pinterest'. 
Interestingly, even the seed node 'Network science' is not on the top ten nodes according to the stationary distribution. 
Here, again, the simple RSP betweenness makes an interesting compromise between these two extremes by respecting the high connectivity of the social networks cluster while also highlighting important, general nodes, such as 'Mathematics' and 'Computer Science' from the peripheral group.

This example indicates the potential of the simple RSP betweenness in analyzing and exploiting semantic and associative networks, as in \cite{kivimaki2013graph}. 
Moreover, the example can help in applications of web design and marketing, for instance, considering situations where a user tries to find a certain page on a web site by following hyperlinks of that site. 
This can happen, for example by browsing videos on \href{http://youtube.com/}{Youtube}, or when playing the \href{http://thewikigame.com/}{Wiki Game}, in which the purpose is to find a target Wikipedia page from a starting page only by clicking the hyperlinks on the pages.

\section{Conclusion}

We have presented two new graph node betweenness centrality measures based on Randomized Shortest Paths. 
The first measure, the simple RSP betweenness centrality, counts the expected number of visits to a node, while the second, the RSP net betweenness, is based on the overall net flow over edges connected to a node. 
Both of these measures are parametrized generalizations of more traditional betweenness centrality measures. 
The RSP net betweenness and its high-temperature limit function, the current flow betweenness seem theoretically more elaborate than the simple RSP betweenness and its limit function, the stationary distribution. 
However, based on our experiments, the simple RSP betweenness seems to provide more satisfying and practical results than the net betweenness, in addition to which (and perhaps partly because) it is easier to interpret. 
In general, our experiments have shown that the RSP betweenness centralities can provide interesting insight into the role and importance of the nodes in a network in ways that the more traditional betweenness measures based on either shortest paths or unbiased random walks can not achieve.

The RSP betweenness measures could be further compared with other centrality measures, which could be a subject for future work. 
However, the main purpose of this paper is only to focus on betweenness centrality measures, to introduce the RSP based methods and their computation and to provide some examples where one could benefit from using them. 
Also, the paper only considers betweenness as a global measure on nodes, but the methods can easily be extended to other uses such as edge betweenness, betweenness w.r.t. 
a group of nodes or betweenness between groups of source and target nodes, all of which have relevant applications. 
One drawback of the RSP framework is that it often requires a full matrix inverse, because of which it is not currently practical for very large networks. 
One topic for future research is to develop methods that would allow estimating the RSP-based quantities for large networks either by using more specialized computational methods or by approximation.

Although the computational complexity of the RSP-based methods can be too high for very large problems, the implementation and the interpretation of the computations is quite straightforward. 
Moreover, the framework lies on solid theoretical grounds, and considering the generalization of shortest paths by randomization makes sense for many application scenarios. 
In our examples we have shown them to give promising results in highlighting nodes that belong to a central group of nodes, in detecting possible bottlenecks in street networks for navigation modeling and also in evaluating the visit rate of pages on the World Wide Web.

\begin{acknowledgments}
This work has been funded by Emil Aaltonen Foundation, InnovIris and the R\'egion Wallonne. 
I.K.\ and J.S.\ also acknowledge support from the Academy of Finland, project no.\ 260427.
I.K.\ would like to thank Jean-Charles Delvenne and Rainer Kujala for useful discussions and advice and Dries Verdegem for providing the parsed Wikipedia hyperlink network.
\end{acknowledgments}

\bibliography{RSPBC_biblio_arxiv}

\end{document}